\documentclass[prl,twocolumn,a4]{revtex4}
\usepackage{graphicx}
\usepackage{epsfig}
\usepackage{amssymb,latexsym,amsmath}
\newcommand{\bs}{\mathbf}
\usepackage{hyperref}

\usepackage{ulem}
\usepackage{color}

\begin{document}
\title{ \Large{ \bf A new approach to time-dependent transport through an interacting quantum dot within Keldysh formalism } }

\author{V. Vovchenko$^a$, D. Anchishkin$^b$, J. Azema$^c$, P. Lombardo$^c$, R. Hayn$^c$, A.-M. Dar\'{e}$^c$\email{Anne-Marie.Dare@univ-amu.fr}
\\ $^a${\small \it Taras Shevchenko National University of Kiev,
03022 Kiev, Ukraine}
\\ $^b${\small \it Bogolyubov Institute for Theoretical Physics,
             03680 Kiev, Ukraine}
\\ $^c${\small \it Aix-Marseille Universit\'e, CNRS, IM2NP UMR 7334,}{ \small \it  13397, Marseille, France}}

\date{\today}

\begin{abstract}
The time-dependent transport through a nano-scale device, consisting
of a single spin-degenerate orbital with on-site Coulomb interaction, coupled to
two leads, is investigated.
Various gate and bias voltage time-dependences are considered.
The key and new point lies in the proposed way to avoid the difficulties of the usual heavy computation when dealing with two time Green's functions within Keldysh formalism.
The time-dependent retarded dot Green's functions
are evaluated,
in an efficient manner within a non-canonical
Hubbard I approximation.
Calculations
of the time-dependent current are then presented in the wide-band limit
for different parameter sets.
A comparison between the method and the Hartree-Fock approximation is performed as well.
It is shown that the later cannot account reliably for dynamical aspects of transport phenomena.
\end{abstract}
\maketitle

\section{Introduction}
The investigation of electron transport in nano-structures, such as quantum dots or carbon
nanotubes, is of high actual interest. It
leads
to the observation of a multitude of mesoscopic phenomena,
where the dual nature of quasi-particles is readily observed. Its wave character is manifest in
interference phenomena due to the phase coherence of charge carriers, whereas its granular character
is best visible in the Coulomb blockade
effect \cite{Altshuler91,DattaBook,HaugJauhoBook}. Besides their fundamental character,
mesoscopic transport studies are also of high technological interest. It is sufficient to quote the development
of carbon nanotube transistors \cite{Appenzeller04}, which can operate up to
terahertz frequencies \cite{Zhong08}.

Several complications arise for the theoretical description of transport properties in nanodevices
which are
beyond solid state physics outlined in classical text-books.
These include nonadiabatic effects
and time dependent
phenomena,
as well as
the wealth of properties induced by
Coulomb correlations. For a realistic description,
all these effects have to be treated simultaneously.
This is a real challenge.
Nonadiabatic time dependent effects are usually treated
within the framework of two-time Keldysh Green's functions \cite{JauhoWingreenMeir1994}, but the treatment
of the Coulomb interaction is indispensable in most cases. One can distinguish three levels of correlation
effects: (i) weak Coulomb correlation of either Hartree (electrostatic interaction) or Hartree-Fock (including
spin exchange) character, (ii) Coulomb blockade, and (iii) Kondo physics. The two crucial parameters that
guide Coulomb effects are the coupling between dot and electrodes, and the temperature. By decreasing
these two parameters the Coulomb effects become more and more important,
the Kondo physics taking place at low temperature, for a low bias voltage.
To simulate the terahertz response of carbon nanotube transistors, the treatment of
weak correlation effects (in Hartree approximation) is already a standard procedure \cite{Kienle09}.
However, a well developed
time-dependent formalism
in case of strong correlation and a fortiori when Coulomb-induced collective phenomena set in,
deserves to be improved, and is currently a very active research area.

Here, we present a formalism of non-adiabatic electron transport which is able to treat an arbitrary
time-dependence
of bias and gate voltages, as well as a time-dependence of the hybridization between leads and dot.
The transient and steady-state properties can be evaluated, without any a priory assumption about adiabatic or sudden limits.
Our formalism includes the effects of Coulomb
correlations beyond weak-coupling treatment, and works
for arbitrary on-site interaction $U$.
Our approach is not restricted to the wide-band limit, although the presented numerical results rely on it.
It treats correctly the uncorrelated ($U \to 0$) and the atomic (disconnected dot) limits, but cannot account for the Kondo effect,
albeit interplay between transient phenomena and Kondo physics are very interesting issues~\cite{Nordlander99},~\cite{PlihalLangreth2000}.
As a consequence the temperatures considered here are higher than the Kondo temperature $T_K$.

The formalism is based on a systematic use of two-time Keldysh Green's functions in the Hubbard I
approximation (HIA)~\cite{Hubbard1963,Hewson66}.
The key and new point lies in the proposed way to circumvent the difficulties of the usual heavy computation when dealing with two time Green's functions.
Our approach is easy to implement and constitutes a substantial computer-time saving method.
This advantage is shared with the technique recently developed by Croy and Saalman \cite{CroySaalmann85}.
The approaches are different in detail but predict consistent results, as shown later.

After having presented the general formalism in Sec. 2, we show in some detail
the main idea of our approximation in Sec. 3.
First we apply the approach to a steady-state case (Sec.4). This allows us to compare our correlation treatment to the Hartree-Fock (HF) approximation, and also to a more sophisticated correlation treatment, namely the noncrossing approximation (NCA)~\cite{Bickers}. For the parameters under study, the NCA 
 with a renormalized Hubbard bandwidth
and 
 HIA in the non-equilibrium steady-state regime compare favorably,
while HF turns to be unreliable~\cite{myohaetal}.
In the time-dependent situation (Sec. 5) we investigate the case of a pulse modulation for the bias voltage, which enables to measure  the charging time in the Coulomb blockade regime, as well as the tunnel and displacement currents.
We then investigate the case of a forced harmonic bias voltage, with transient and steady-state regimes, before turning to the case of a pure pumping experiment, where, along the lines developed by Croy {\it et al.}~\cite{CroySaalmann85}, we address the question of adiabatic~\cite{Splettstoesser05} and
non-adiabatic frontier. We finally close with our Conclusions (Sec.\ 6).

\section{Model Hamiltonian and general expression for time-dependent current}

We consider a system which is a single level interacting Anderson quantum dot, coupled
to two uncorrelated leads, acting as source and drain.
The Hamiltonian reads~\cite{Anderson}
\begin{equation}
H = H_c + H_T + H_{\rm cen}.
\end{equation}
$H_c$ is a contact Hamiltonian corresponding to electrons in leads, and takes the
following form
\begin{equation}
H_c = \sum_{k \alpha \sigma} \varepsilon_{k \alpha} (t) \bs c_{k \alpha \sigma}^\dag \bs c_{k \alpha \sigma},
\end{equation}
where $k$ denotes
momentum index,
$\alpha = (L,R)$ stands for left and right leads and
$\sigma = (\uparrow, \downarrow)$
is the spin degree of freedom.
Here $\bs c_{k \alpha \sigma}^\dag$ and $\bs c_{k \alpha \sigma}$ are electron creation and
annihilation operators for the $\alpha$-lead state $k \sigma$.
$H_T$ is a coupling term
\begin{equation}
H_T = \sum_{k \alpha \sigma} \left[ V_{k \alpha} (t) \bs c_{k \alpha \sigma}^\dag \bs d_{\sigma} +
V_{k \alpha}^* (t) \bs d_{\sigma}^\dag \bs c_{k \alpha \sigma}\right],
\end{equation}
where $\bs d_{\sigma}^\dag$ and $\bs d_{\sigma}$ are electron creation and
annihilation operators at the dot for spin state $\sigma$.
The central region Hamiltonian $H_{\rm cen}$ includes Coulomb repulsion term:
\begin{equation}
H_{\rm cen} = \varepsilon_0 (t) \sum_{\sigma} \bs d_{\sigma}^\dag \bs d_{\sigma} + U \bs d_{\uparrow}^\dag \bs d_{\uparrow} \bs d_{\downarrow}^\dag \bs d_{\downarrow}.
\end{equation}
The energy level in the dot $\varepsilon_0(t)$, in the leads
$\varepsilon_{k \alpha} (t)= \varepsilon_{k}^0 + \Delta_{\alpha} (t)$,
as well as the hybridization coefficients  $V_{k \alpha}(t)$,
are all considered to be time dependent, and independent of each other. Experimentally, that can be
realized by applying different bias and gate voltages.
We shall use the assumption that the time dependence of the
hybridization parameters can be factorized as
$V_{k \alpha} (t)=u_{\alpha}(t) V_{\alpha}(\varepsilon^0_{k })$~\cite{JauhoWingreenMeir1994}.

Current from left contact to the central region can be calculated as
\begin{equation}
J_L(t) = -e \langle \dot{N_L} (t) \rangle = - i \frac{e}{\hbar} \langle [H, N_L] \rangle,
\end{equation}
where $N_L$ is the lead fermion number operator.
Similar problem was considered recently \cite{CroySaalmann85} in the wide-band limit
with a density matrix approach using a truncated equation of motion technique.
In the present paper we apply the time-dependent Keldysh formalism~\cite{Keldysh}.
The expression for the current can be written in terms of central region Keldysh Green's functions (see Ref.~\cite{JauhoWingreenMeir1994})
\begin{eqnarray}
J_L(t) & = & -2 \frac{e}{\hbar} \int_{-\infty}^t d t_1 \int \frac{d \varepsilon}{2\pi} \mathrm{Im} \sum_{\sigma}
\left\{  \right. e^{-i \varepsilon (t_1-t)} \Gamma^L (\varepsilon, t_1, t) \nonumber \\
& \times & \left[ G_{\sigma \sigma}^{<} (t,t_1) +
f_L(\varepsilon) G_{\sigma \sigma}^r (t,t_1)  \right] \left. \right\}.
\label{eq:lcur}
\end{eqnarray}
Here $\Gamma^L (\varepsilon, t_1, t)$ is defined as
\begin{eqnarray}
\Gamma^{\alpha} (\varepsilon, t_1, t) & = &  2 \pi \rho(\varepsilon) u_{\alpha}(t) u_{\alpha}(t_1)
V_{\alpha} (\varepsilon)
V_{\alpha}^* (\varepsilon) \nonumber \\
& & \times \exp \left[i \int_{t_1}^t d t_2 \Delta_{\alpha} (t_2) \right],
\end{eqnarray}
where $\rho(\varepsilon)$ is the density of states
per lead and per spin, which we choose independent of $\alpha$.
$f_L(\varepsilon)$ is the Fermi distribution function of the left contact.
Finally $G_{\sigma \sigma}^{<} (t,t_1)$ and  $G_{\sigma \sigma}^r (t,t_1)$ are the lesser and retarded Keldysh Green's functions   of the central region
\begin{eqnarray}
G_{\sigma \sigma}^{<} (t,t') & = & i \langle \bs d_{\sigma}^\dag (t') \bs d_{\sigma} (t) \rangle, \\
G_{\sigma \sigma}^{r} (t,t') & = & -i \operatorname{\theta} (t - t') \langle \{ \bs d_{\sigma} (t), \bs d_{\sigma}^\dag (t')\} \rangle.
\end{eqnarray}
Even out of equilibrium, the Green's functions are diagonal in spin,
due to our choice for $H$ which conserves spin.
In order to calculate the time-dependent current one needs to calculate these Green's functions first.

\section{Green's functions of the central region}
\subsection{Equation of motion for Green's functions}
We calculate the equation of motion for the
retarded Green's function $G_{\sigma \sigma}^{r} (t,t')$.
It leads to
\begin{eqnarray}
\left( i \frac{\partial }{\partial t} -  \varepsilon_{0} (t) \right) G_{\sigma \sigma}^{r} (t,t')  & = & \delta(t-t')
+ \sum_{k \alpha} V^*_{k \alpha}(t) G_{k \alpha \sigma, \sigma}^{r}(t,t') \nonumber \\
& + & U G_{\sigma\sigma,U}^{r} (t,t'),
\label{eq:eqmoGr}
\end{eqnarray}
where two other Green's functions appear: $G_{\sigma\sigma,U}^{r} (t,t') = -i \operatorname\theta(t-t') \langle \{ (\bs d_{\sigma} \bs d_{\bar{\sigma}}^{\dag} \bs d_{\bar{\sigma}}) (t), \bs d_{\sigma}^\dag (t')\} \rangle$
and $G_{k \alpha \sigma , \sigma}^{r} (t,t') = -i \operatorname\theta(t-t') \langle \{ \bs c_{k \alpha \sigma} (t), \bs d_{\sigma}^\dag (t')\} \rangle$.
The equation of motion for $G_{k \alpha \sigma, \sigma}^{r}(t,t')$ is
\begin{equation}
\left(i \frac{\partial }{\partial t} - \varepsilon_{k \alpha} (t) \right) G_{k \alpha \sigma, \sigma}^{r} (t,t') =
V_{k \alpha}(t) G_{\sigma \sigma}^{r}(t,t').
\end{equation}
The formal solution of this equation can be written as
\begin{equation}
G_{k \alpha \sigma, \sigma}^{r} (t,t') = \int d t_1 g_{k \alpha}^{r} (t, t_1) V_{k \alpha}(t_1)
G_{\sigma \sigma}^{r} (t_1, t'),
\label{eq:Gkas}
\end{equation}
where $g_{k \alpha}^{r} (t, t_1)$ is  the Green's function for the uncoupled system
\begin{equation}
g_{k \alpha}^r (t,t') = -i \operatorname{\theta}(t-t') \exp\left[ -i \int_{t'}^t d t_1 \varepsilon_{k \alpha} (t_1) \right].
\end{equation}
Substituting \eqref{eq:Gkas} into the equation for $G_{\sigma \sigma}^{r} (t,t')$ \eqref{eq:eqmoGr}, we get
\begin{eqnarray}
& & \left( i \frac{\partial }{\partial t}  - \varepsilon_{0} (t) \right)  G_{\sigma \sigma}^{r} (t,t')   =  \delta(t-t')  \nonumber \\
&& + \int d t_1 \Sigma^{r} (t, t_1) G_{\sigma \sigma}^{r}(t_1,t')
+ U G_{\sigma\sigma,U}^{r} (t,t'),
\label{eq:Grdt}
\end{eqnarray}
where $\Sigma^{r} (t, t')$ is the hybridization self-energy
\begin{eqnarray}
\Sigma^{r} (t, t') & = &  \sum_{k \alpha} V^*_{k \alpha} (t) g_{k \alpha}^{r} (t, t') V_{k \alpha} (t') \nonumber \\
& = & -i \operatorname{\theta}(t-t') \sum_{\alpha} \int \frac{d \varepsilon}{2\pi} e^{-i \varepsilon (t-t')}
\Gamma^{
\alpha
}(\varepsilon, t, t'). \nonumber\\
& & 
\label{eq:sigmar}
\end{eqnarray}
Equation \eqref{eq:Grdt} is not a closed equation for $G_{\sigma \sigma}^{r} (t,t')$ because of the presence of
so far
unknown $G_{\sigma\sigma,U}^{r} (t,t')$.
In order to get a closed equation we need to make certain approximations regarding
this last Green's function.

\subsection{Approximations}

\subsubsection{Hartree-Fock approximation}
Within the HF approximation we use the following factorization~\cite{Hubbard1963,Hewson66}
\begin{eqnarray}
G_{\sigma\sigma,U}^{r} (t,t')  & \approx & -i \operatorname\theta(t-t') n_{\bar{\sigma}}(t) \langle \{ \bs d_{\sigma} (t), \bs d_{\sigma}^\dag (t')\} \rangle \nonumber \\
& = & n_{\bar{\sigma}}(t) G_{\sigma \sigma}^{r} (t,t'),
\end{eqnarray}
where $n_{\bar{\sigma}}(t) = \langle (\bs d_{\bar{\sigma}}^{\dag} \bs d_{\bar{\sigma}}) (t) \rangle$.
In this case we can get an equation for $G_{\sigma \sigma}^{r} (t,t')$
\begin{eqnarray}
& & \left( i \frac{\partial }{\partial t} -  \varepsilon_{0} (t) - n_{\bar{\sigma}}(t)U \right) G_{\sigma \sigma}^{r} (t,t') = \delta(t-t') \nonumber \\
& & \qquad + \int d t_1 \Sigma^{r} (t, t_1) G_{\sigma \sigma}^{r}(t_1,t')).
\label{eq:GrHF}
\end{eqnarray}
The quantity $n_{\bar{\sigma}}(t)$ can be determined from the lesser Green's function
as $n_{\bar{\sigma}}(t) =
 \text{\rm Im} G_{\bar{\sigma} \bar{\sigma}}^< (t,t)
$ and therefore
a self-consistent scheme is needed to solve eq.~\eqref{eq:GrHF}.

In the HF approximation, the retarded dot Green's function
has the same form as in the case of non-correlated electrons, the only
difference being $\varepsilon_0(t) \to \varepsilon_0(t) + n_{\bar{\sigma}}(t)U$.
It means that the Hartree-Fock approximation reduces here to the Hartree approximation, which is quite crude, as will be seen later.

\subsubsection{ Non-canonical Hubbard I approximation}
It is possible to get a better approximation for $G_{\sigma\sigma,U}^{r} (t,t')$
by considering the equation of motion for this function. First we write down
the commutator $[\bs d_{\sigma} \bs d_{\bar{\sigma}}^{\dag} \bs d_{\bar{\sigma}}, \bs H]$
and neglect all the terms
without coinciding quantum numbers for at least two fermionic operators:
\begin{eqnarray}
&& [\bs d_{\sigma} \bs d_{\bar{\sigma}}^{\dag} \bs d_{\bar{\sigma}}, \bs H]  =  (\varepsilon_0 (t) + U)\bs d_{\sigma} \bs d_{\bar{\sigma}}^{\dag} \bs d_{\bar{\sigma}} \nonumber \\
&&+ \sum_{k \alpha} \left( V_{k \alpha}(t) \bs c_{k \alpha \bar{\sigma}}^{\dag} \bs d_{\sigma} \bs d_{\bar{\sigma}} +
V^*_{k \alpha}(t) \bs d_{\bar{\sigma}}^{\dag} \bs c_{k \alpha \bar{\sigma}} \bs d_{\sigma}
+ V^*_{k \alpha}(t) \bs c_{k \alpha \sigma} \bs d_{\bar{\sigma}}^{\dag} \bs d_{\bar{\sigma}} \right)  \nonumber \\
 && \qquad \approx   (\varepsilon_0 (t) + U)  \bs d_{\sigma} \bs d_{\bar{\sigma}}^{\dag} \bs d_{\bar{\sigma}} + \sum_{k \alpha} V^*_{k \alpha}(t) \bs c_{k \alpha \sigma} \bs d_{\bar{\sigma}}^{\dag} \bs d_{\bar{\sigma}}.
\label{eq:Comm}
\end{eqnarray}
At this point we could follow the usual HIA procedure (henceforth called canonical), with the factorization $\bs c_{k \alpha \sigma} \bs d_{\bar{\sigma}}^{\dag} \bs d_{\bar{\sigma}} \approx  \bs c_{k \alpha \sigma}
\langle \bs d_{\bar{\sigma}}^{\dag} \bs d_{\bar{\sigma}} \rangle $, as used for example in Ref.~\cite{Hewson66}. This would close the system of equations. However this decoupling scheme leads to cumbersome numerical difficulties, while keeping this term untouched as shown later, enables to skirt them.
The equation of motion for $G_{\sigma\sigma',U}^{r} (t,t')$ then takes the following form
\begin{eqnarray}
\left( i \frac{\partial }{\partial t} -  \varepsilon_{0} (t) - U \right) G_{\sigma\sigma,U}^{r} (t,t')
&= &\delta(t-t') n_{\bar{\sigma}} (t) \nonumber \\
&+& \sum_{k \alpha} V^*_{k \alpha} (t) G_{k \alpha \sigma, \sigma, U}^{r} (t,t'),\nonumber \\
&&
\label{eq:systemAdd1}
\end{eqnarray}
where appears a new Green's function
$G_{k \alpha \sigma, \sigma, U}^{r} (t,t') = -i \operatorname\theta(t-t') \langle \{ (\bs c_{k \alpha \sigma} \bs d_{\bar{\sigma}}^{\dag} \bs d_{\bar{\sigma}}) (t), \bs d_{\sigma}^\dag (t')\} \rangle$,
for which, neglecting the same kind of terms as in \eqref{eq:Comm}, the equation of motion leads to
\begin{equation}
\left( i \frac{\partial }{\partial t} -  \varepsilon_{k \alpha} (t)\right) G_{k \alpha \sigma, \sigma, U}^{r} (t,t')
= V_{k \alpha}(t) G_{\sigma\sigma,U}^{r} (t,t').
\label{eq:systemAdd2}
\end{equation}
Eqs. \eqref{eq:systemAdd1} and \eqref{eq:systemAdd2} form a
closed
set of equations
for $G_{\sigma\sigma,U}^{r} (t,t')$ and $G_{k \alpha \sigma, \sigma, U}^{r} (t,t')$.
It is useful to rewrite this system
in integral form.
To do so we define the following functions
\begin{eqnarray}
g_{0}^r (t,t') & = & -i \operatorname{\theta}(t-t') \exp\left[ -i \int_{t'}^t d t_1 \varepsilon_{0} (t_1) \right], \label{eq:g1}\\
g_{U}^r (t,t') & = & -i \operatorname{\theta}(t-t') \exp\left[ -i \int_{t'}^t d t_1 \varepsilon_{0} (t_1) \right] \nonumber \\
& \quad & \times  \exp\left[-iU(t-t')\right],\label{eq:g2}\\
\tilde{g}_{0}^r (t,t') & = & (1 - n_{\bar{\sigma}} (t')) g_{0}^r (t,t'),\label{eq:g3}\\
\tilde{g}_{U}^r (t,t') & = & n_{\bar{\sigma}} (t') g_{U}^r (t,t'),
\label{eq:g4}
\end{eqnarray}
and rewrite the equations \eqref{eq:systemAdd1} and \eqref{eq:systemAdd2}
\begin{eqnarray}
G_{\sigma\sigma,U}^{r} (t,t') & = & \tilde{g}_{U}^r (t,t')  \nonumber \\
&+& \int d t_1 g_{U}^r (t,t_1) \sum_{k \alpha}
V^*_{k \alpha} (t_1) G_{k \alpha \sigma, \sigma, U}^{r} (t_1,t')
\nonumber  \\
&& \label{eq:intAdd1} \\
G_{k \alpha \sigma, \sigma, U}^{r} (t,t') & = & \int d t_1 g_{k \alpha}^r (t,t_1) V_{k \alpha} (t_1)
G_{\sigma\sigma,U}^{r} (t_1,t').
\label{eq:intAdd2}
\end{eqnarray}
Substituting Eq.~\eqref{eq:intAdd2} into \eqref{eq:intAdd1} and using \eqref{eq:sigmar} for $\Sigma^{r}_{\sigma}$,
we get an
integral equation for $G_{\sigma\sigma,U}^{r} (t,t')$ only
\begin{eqnarray}
G_{\sigma\sigma,U}^{r} (t,t') & = & \tilde{g}_{U}^r (t,t') \nonumber \\
&+& \int d t_1 d t_2 g_{U}^r (t,t_1) \Sigma^r(t_1, t_2) G_{\sigma \sigma, U}^{r} (t_2,t').
\nonumber \\
& & \label{eq:GUD}
\end{eqnarray}

Going back to the integral equation for the local Green's function resulting from \eqref{eq:Grdt}
\begin{eqnarray}
G_{\sigma\sigma}^{r} (t,t') & = & g_{0}^{r} (t,t') + \int d t_1 d t_2 g_{0}^{r}(t,t_1) \Sigma^{r} (t_1,t_2) G_{\sigma\sigma}^{r} (t_2, t') \nonumber \\
& + & U \int d t_1 g_{0}^{r} (t, t_1) G_{\sigma\sigma,U}^{r} (t_1,t')
\label{eq:Grint}
\end{eqnarray}
and using \eqref{eq:GUD}, we obtain
\begin{eqnarray}
G_{\sigma\sigma}^{r} (t,t') & = & g_{0}^{r} (t,t') + \tilde{g}_{U}^{r} (t, t') - n_{\bar{\sigma}}(t')g_0^{r} (t, t')  \nonumber \\
& + &  \int d t_1 d t_2 g_{0}^{r}(t,t_1) \Sigma^{r} (t_1,t_2) G_{\sigma\sigma}^{r} (t_2, t') \nonumber \\
&+ &  \int d t_1 d t_2 \left(g_{U}^{r} (t, t_1) - g_{0}^{r} (t, t_1) \right)  \nonumber \\
& \times &\Sigma^{r} (t_1, t_2) G_{\sigma\sigma,U}^{r} (t_2,t').
\end{eqnarray}
Next we will decompose the central region Green's function into two components
$G_{\sigma\sigma}^{r} (t,t') = G_{\sigma\sigma,0}^{r} (t,t') + G_{\sigma\sigma,U}^{r} (t,t')$,
where $G_{\sigma\sigma,U}^{r} (t,t')$ satisfies the equation \eqref{eq:GUD} and $G_{\sigma\sigma,0}^{r} (t,t')$
is a new unknown function. Then,
using the definition for $\tilde{g}_{0}^{r} (t,t')$
we can write
\begin{eqnarray}
G_{\sigma\sigma,0}^{r} (t,t') &+ &G_{\sigma\sigma,U}^{r} (t,t')  =  \tilde{g}_{0}^{r} (t,t') + \ \tilde{g}_{U}^{r} (t, t') \nonumber \\
&+ &\int d t_1 d t_2 g_{0}^{r}(t,t_1) \Sigma^{r} (t_1,t_2) G_{\sigma\sigma,0}^{r} (t_2, t')  \nonumber \\
&+ &  \int d t_1 d t_2 g_{U}^{r} (t, t_1)\Sigma^{r} (t_1, t_2) G_{\sigma\sigma,U}^{r} (t_2,t'). \nonumber \\
& &
\end{eqnarray}
It is seen
from \eqref{eq:GUD} that the last two terms here add up to form $G_{\sigma\sigma,U}^{r} (t,t')$.
Thus we can get the integral equation for $G_{\sigma\sigma,0}^{r} (t,t')$
\begin{equation}
G_{\sigma\sigma,0}^{r} (t,t') = \tilde{g}_{0}^{r} (t,t') + \int d t_1 d t_2 g_{0}^{r}(t,t_1) \Sigma^{r} (t_1,t_2) G_{\sigma\sigma,0}^{r} (t_2, t').
\label{eq:G0D}
\end{equation}
To summarize
we have decomposed the diagonal central region Green's function
into two terms $G_{\sigma\sigma}^{r} (t,t') = G_{\sigma\sigma,0}^{r} (t,t') + G_{\sigma\sigma,U}^{r} (t,t')$, where
$G_{\sigma\sigma,0}^{r} (t,t')$ and $G_{\sigma\sigma,U}^{r} (t,t')$ satisfy equations \eqref{eq:G0D} and \eqref{eq:GUD}
respectively.
The terms $G_{\sigma\sigma,0}^{r}$ and $G_{\sigma\sigma,U}^{r}$  actually correspond to the lower and upper Hubbard bands
as will be detailed later.

\subsection{Lesser Green's functions}

To get the lesser Green's functions we will use  the analytic continuation
rules of the Keldysh formalism. First, let us introduce the
Green's functions which are related to the Hubbard bands in the following way
\begin{eqnarray}
\tilde{G}_{\sigma\sigma,0}^{r} (t,t') & = & \frac{G_{\sigma\sigma,0}^{r} (t,t')}{1 - n_{\bar{\sigma}}(t')}\\
\tilde{G}_{\sigma\sigma,U}^{r} (t,t') & = & \frac{G_{\sigma\sigma,U}^{r} (t,t')}{n_{\bar{\sigma}}(t')}
\end{eqnarray}
From Eqs. \eqref{eq:G0D} and \eqref{eq:GUD}
we get the equations for the newly introduced Green's functions
\begin{eqnarray}
\tilde{G}_{\sigma\sigma,0}^{r} (t,t') & = & g_{0}^{r} (t,t') \nonumber \\
&+& \int d t_1 d t_2 g_{0}^{r} (t,t_1) \Sigma^{r} (t_1,t_2) \tilde{G}_{\sigma\sigma,0}^{r} (t_2, t'),\nonumber \\
& & \label{eq:final1} \\
\tilde{G}_{\sigma\sigma,U}^{r} (t,t') & = & g_{U}^{r} (t,t') \nonumber \\
&+& \int d t_1 d t_2 g_{U}^{r} (t,t_1) \Sigma^r(t_1, t_2) \tilde{G}_{\sigma \sigma, U}^{r} (t_2,t'). \nonumber \\
& & \label{eq:final2}
\end{eqnarray}
These equations are actually of the type of
standard Dyson equations for the retarded Green's function of the central region
in the absence of electron interaction at the dot (see Ref.~\cite{JauhoWingreenMeir1994}), the only
difference being the change $\varepsilon_0(t) \to \varepsilon_0(t) + U$ for the
disconnected Green's function in the second equation.
We will introduce $\varepsilon_U(t)$ as $\varepsilon_U(t) = \varepsilon_0(t) + U$ for convenience.
The total retarded Green's function of the central region can then be expressed as
\begin{equation}
G_{\sigma\sigma}^{r} (t,t') = \left(1 - n_{\bar{\sigma}}(t')\right) \tilde{G}_{\sigma\sigma,0}^{r} (t,t') + n_{\bar{\sigma}}(t') \tilde{G}_{\sigma\sigma,U}^{r} (t,t').\\
\label{eq:Grtottilde}
\end{equation}
The last three equations constitute the key results of our HIA.
It will be then straightforward to get the lesser Green's functions and to do numerical calculations,
thanks to the fact that
$\Bigl(  g^r_{0} (t,t') \Bigr)^{-1}$ and $\Bigl(  g^r_{U} (t,t') \Bigr)^{-1}$ depend only on one time variable, namely $t$.

Now we can get the expressions for the corresponding lesser Green's functions which were derived
in Ref. \cite{JauhoWingreenMeir1994} using the Langreth rules
\begin{eqnarray}
\tilde{G}_{\sigma\sigma,0(U)}^{<} (t,t') & = & \int d t_1 d t_2 \tilde{G}_{\sigma\sigma,0(U)}^{r} (t,t_1)
\Sigma^{<} (t_1,t_2) \nonumber \\
& \times & \ \tilde{G}_{\sigma\sigma,0(U)}^{a} (t_2, t'), \label{eq:less12}
\end{eqnarray}
where the lesser self-energy  $\Sigma^< (t, t')$ is
\begin{eqnarray}
\Sigma^< (t, t')& =& \sum_{k \alpha} V_{k \alpha}^* (t) g_{k \alpha}^< (t,t') V_{k \alpha} (t') \nonumber \\
&=& i \sum_{\alpha} \int \frac{d \varepsilon}{2\pi} e^{-i \varepsilon (t-t')} f(\varepsilon) \Gamma^{\alpha}(\varepsilon, t, t')  ,
\end{eqnarray}
and the advanced Green's function is $\tilde{G}_{\sigma\sigma,0(U)}^{a} (t, t') = \left[\tilde{G}_{\sigma\sigma,0(U)}^{r} (t',t)\right]^*$.
The total lesser Green's function of the central region reads
\begin{equation}
G_{\sigma\sigma}^{<} (t,t') = \left(1 - n_{\bar{\sigma}}(t')\right) \tilde{G}_{\sigma\sigma,0}^{<} (t,t') + n_{\bar{\sigma}}(t') \tilde{G}_{\sigma\sigma,U}^{<} (t,t').\\
\end{equation}

\section{Time-independent case}
In the time-independent case the Green's functions can be calculated with the use of Fourier transform.
Let us calculate the
self-energy first, and define useful quantities
\begin{eqnarray}
\Sigma^{r(a)} (\omega) & = & \sum_{k \alpha} \frac{|V_{k \alpha}|^2}{\omega - \varepsilon_{k \alpha} \pm i \eta} \nonumber \\
& = &
(\Lambda^L(\omega) + \Lambda^R(\omega)) \mp \frac{i}{2} ( \Gamma^L(\omega) + \Gamma^R(\omega) )  \nonumber \\
& = & \Lambda(\omega) \mp \frac{i}{2} \Gamma(\omega)\\
\Sigma^{<} (\omega) & = & \sum_{k \alpha} |V_{k \alpha}|^2 g_{k \alpha}^< (\omega)  \nonumber \\
&= &i \left[ \Gamma^L (\omega) f_L(\omega) + \Gamma^R (\omega) f_R(\omega) \right].
\end{eqnarray}
The corresponding
Dyson
equations for central region retarded (advanced) Green's functions
in HIA
are
\begin{eqnarray}
\tilde{G}_{\sigma\sigma,0(U)}^{r(a)} (\omega) & = & g_{0(U)}^{r(a)} (\omega) + g_{0(U)}^{r(a)} (\omega) \Sigma^{r(a)} (\omega) \tilde{G}_{\sigma\sigma,0(U)}^{r(a)} (\omega). \nonumber \\
&&
\end{eqnarray}
It is easy to obtain the explicit form for these Green's functions
\begin{eqnarray}
\tilde{G}_{\sigma\sigma,0(U)}^{r(a)} (\omega) & = & \frac{1}{\omega - \varepsilon_{0(U)} - \Lambda(\omega) \pm \frac{i}{2}\Gamma(\omega)},
\end{eqnarray}
from which we get
the retarded (advanced) Green's function of the central region:
\begin{eqnarray}
G_{\sigma \sigma}^{r(a)} (\omega) &=& \frac{1-n_{\bar\sigma}}{\omega - \varepsilon_0 - \Lambda(\omega) \pm \frac{i}{2}\Gamma(\omega)} \nonumber \\
&+& \frac{n_{\bar\sigma}}{\omega - \varepsilon_0 - U - \Lambda(\omega) \pm \frac{i}{2}\Gamma(\omega)}.
\end{eqnarray}
The spectral function $A_{\sigma}(\omega) = i \left[G_{\sigma \sigma}^{r} (\omega) - G_{\sigma \sigma}^{a} (\omega)\right]$ is then
\begin{eqnarray}
A_{\sigma}(\omega) & = & A_{\sigma}^{0}(\omega) + A_{\sigma}^{U}(\omega) \nonumber \\
& = &\frac{(1-n_{\bar\sigma}) \Gamma(\omega)}{\left[\omega - \varepsilon_0 - \Lambda(\omega)\right]^2 + \left[\frac{\Gamma(\omega)}{2} \right]^2} \nonumber \\
& & + \frac{n_{\bar\sigma} \Gamma(\omega)}{\left[\omega - \varepsilon_0 - U - \Lambda(\omega)\right]^2 + \left[\frac{\Gamma(\omega)}{2} \right]^2}.
\label{eq:spectral}
\end{eqnarray}
One recovers the exact results in the atomic limit $V_{k \alpha} (t) \rightarrow 0 $, and in the noninteracting $U \rightarrow 0$ one.
It is straightforward to calculate the lesser Green's function as well
\begin{eqnarray}
G_{\sigma\sigma}^{<} (\omega)
& = & i \left[ \Gamma^L (\omega) f_L(\omega) + \Gamma^R (\omega) f_R(\omega) \right] \nonumber \\
& \times & \Bigl(   \frac{1-n_{\bar{\sigma}}}{\left[\omega - \varepsilon_0 - \Lambda(\omega)\right]^2 + \left[\frac{\Gamma(\omega)}{2} \right]^2}  \nonumber \\
& & + \frac{n_{\bar{\sigma}}}{\left[\omega - \varepsilon_0 - U - \Lambda(\omega)\right]^2 + \left[ \frac{\Gamma(\omega)}{2} \right]^2} \Bigr)
\end{eqnarray}
In the time-independent case it is possible to perform the $t_1$ integration in Eq.~\eqref{eq:lcur} and to
express the current
as  \cite{JauhoWingreenMeir1994}
\begin{eqnarray}
J_{L(R)} &=& i \frac{e}{\hbar} \int \frac{d \varepsilon}{2\pi} \sum_{\sigma} \left\{ \Gamma^{L(R)} (\varepsilon) \left[ G_{\sigma \sigma}^{<} (\varepsilon) \right. \right.\nonumber \\
&&+\left. \left. f_{L(R)}(\varepsilon) \left( G_{\sigma \sigma}^r (\varepsilon) - G_{\sigma \sigma}^a (\varepsilon)  \right) \right] \right\}.
\end{eqnarray}
In terms of spectral function, this leads to
\begin{eqnarray}
J_{L(R)} &=& -\frac{e}{\hbar} \int \frac{d \varepsilon}{2\pi} \sum_{\sigma} A_{\sigma}(\varepsilon)  \frac{\Gamma^{L} (\varepsilon) \Gamma^{R} (\varepsilon)}{\Gamma (\varepsilon)} \nonumber \\
&& \times \left[ f_{R(L)} (\varepsilon) - f_{L(R)} (\varepsilon) \right],
\label{eq:JLReq}
\end{eqnarray}
while the expression for
dot occupancy  for spin $\sigma$
is
\begin{equation}
n_{\sigma} = \mathrm{Im} G_{\sigma\sigma}^{<} (t, t) = \int \frac{d \varepsilon}{2\pi} \,
\displaystyle \bar{f} (\varepsilon) \, A_{\sigma} (\varepsilon),
\label{eq:nequigen}
\end{equation}
where $\displaystyle \bar{f} (\varepsilon) = \frac{\Gamma^L (\varepsilon) f_L(\varepsilon) + \Gamma^R (\varepsilon) f_R(\varepsilon)}{\Gamma(\varepsilon)}$.
Here $A_{\sigma}$ depends on $n_{\bar{\sigma}}$, therefore \eqref{eq:nequigen}
is implicit. We can also write this equation in the following form
\begin{equation}
n_{\sigma} = (1-n_{\bar{\sigma}}) n^0 + n_{\bar{\sigma}} n^U,
\label{eq:nequigen2}
\end{equation}
where
\begin{eqnarray}
n^{0(U)} & = & \int \frac{d \varepsilon}{2\pi} \,
\displaystyle \bar{f} (\varepsilon)
\, \frac{\Gamma(\varepsilon)}{\left[\varepsilon - \varepsilon_{0(U)} - \Lambda(\varepsilon)\right]^2 + \left[\frac{\Gamma(\varepsilon)}{2} \right]^2}.\nonumber \\
&&
\end{eqnarray}
Equation \eqref{eq:nequigen2}, which is actually a system of two linear equations for $n_{\sigma}$ and $n_{\bar{\sigma}}$, can
be solved explicitly to yield
\begin{equation}
n_{\sigma} = n_{\bar{\sigma}} = \frac{n^0}{1+n^0 - n^U}.
\end{equation}

\begin{figure}[h!]
   \begin{center}
     \includegraphics[width=.5\textwidth]{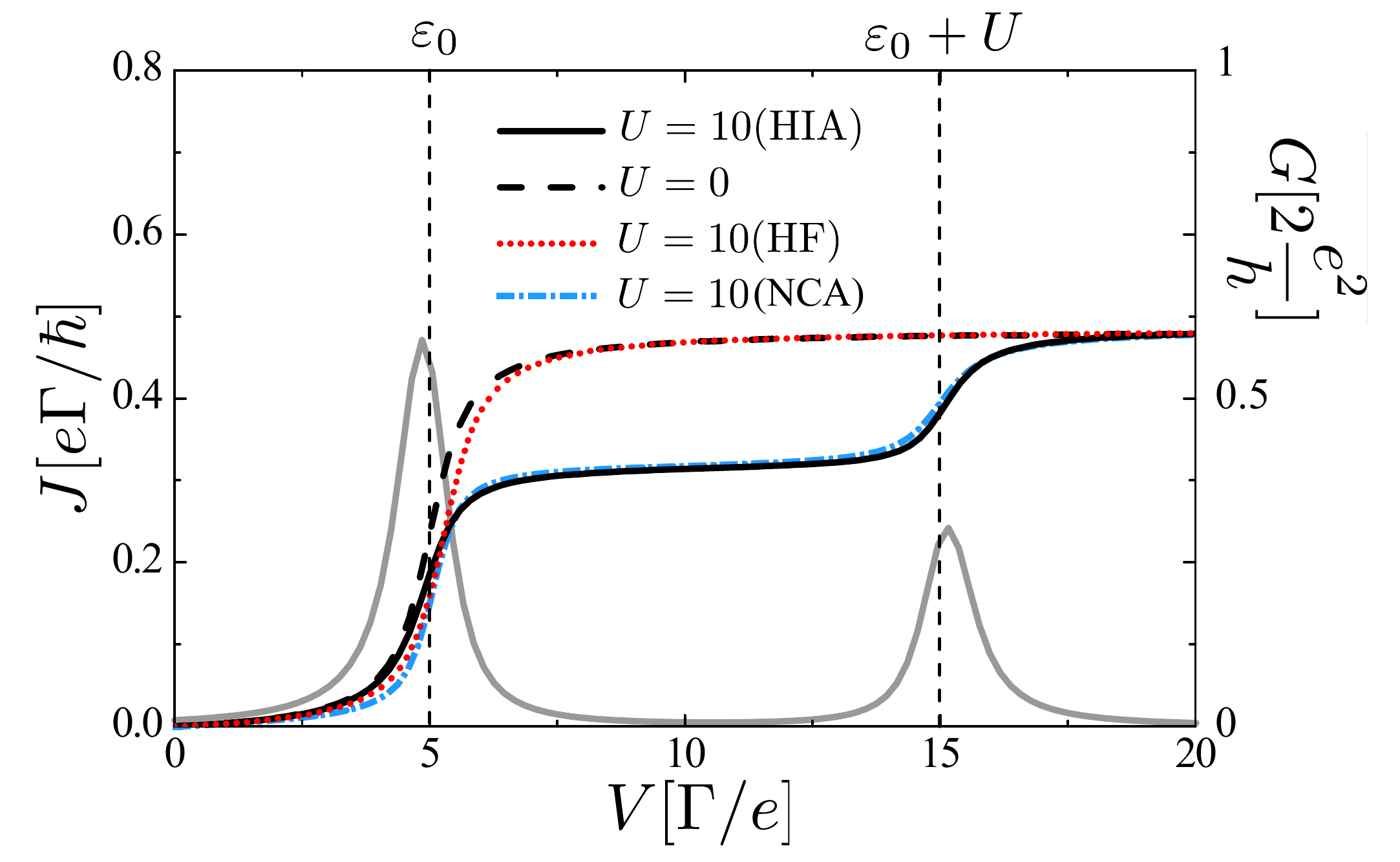}
     \caption{
     Current versus voltage for $\mu_R = 0,
     \mu_L = V,
     \,\varepsilon_{0} = 5,\,U = 10,\,T=0.1,\,\Gamma_L=\Gamma_R=0.5$,
     in the Hubbard I approximation, 
     non interacting case, Hartree-Fock and noncrossing approximations (see text).
    The differential conductance (y axis on the right side) for HIA is also shown in gray.
     }
     \label{pic:VolCur}
   \end{center}
\end{figure}

\begin{figure}[h!]
\begin{minipage}[t]{.48\textwidth}
        \begin{center}
            \includegraphics[width=\textwidth]{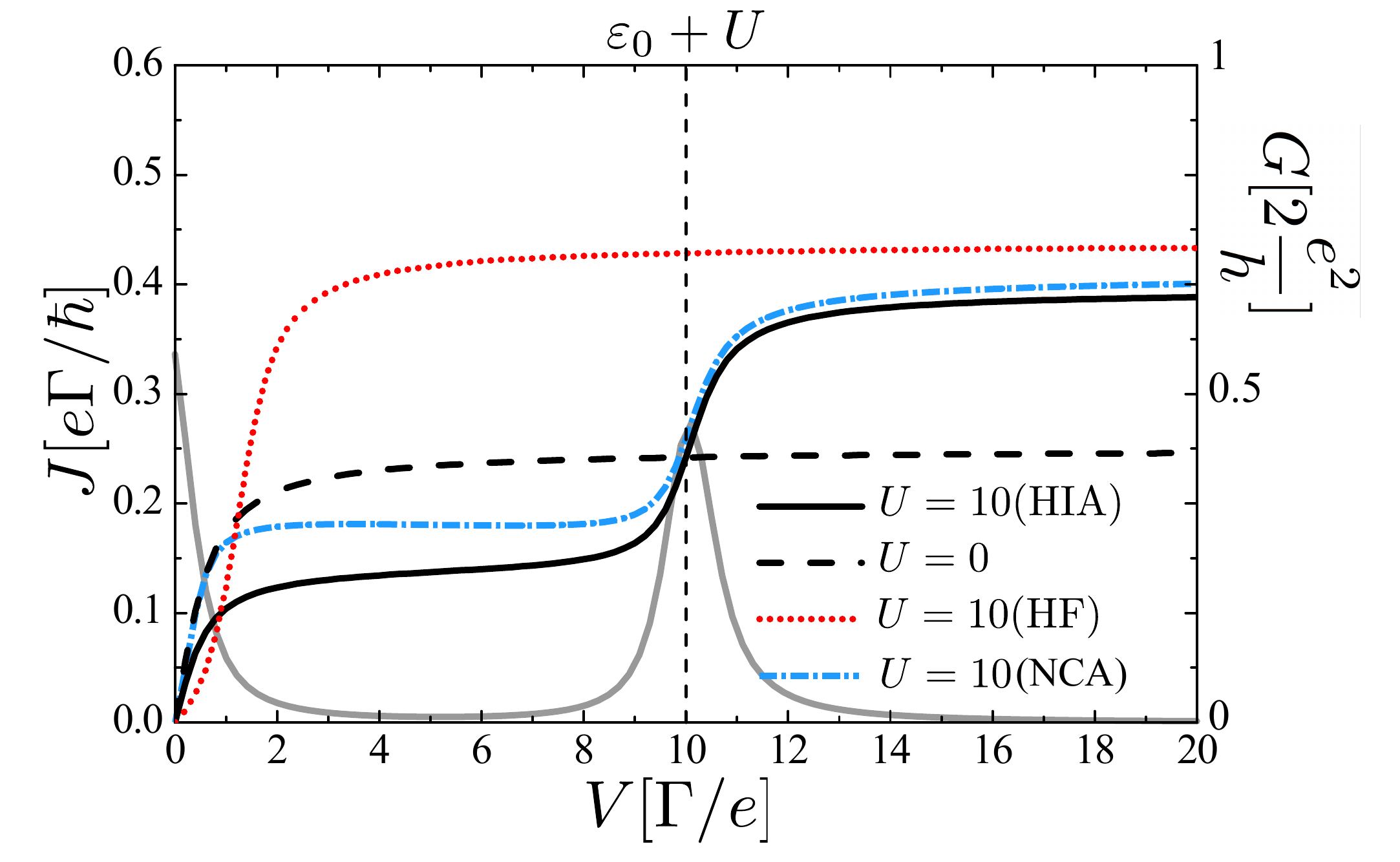}
        \end{center}
    \end{minipage}
    \begin{minipage}[t]{.48\textwidth}
        \begin{center}
            \includegraphics[width=\textwidth]{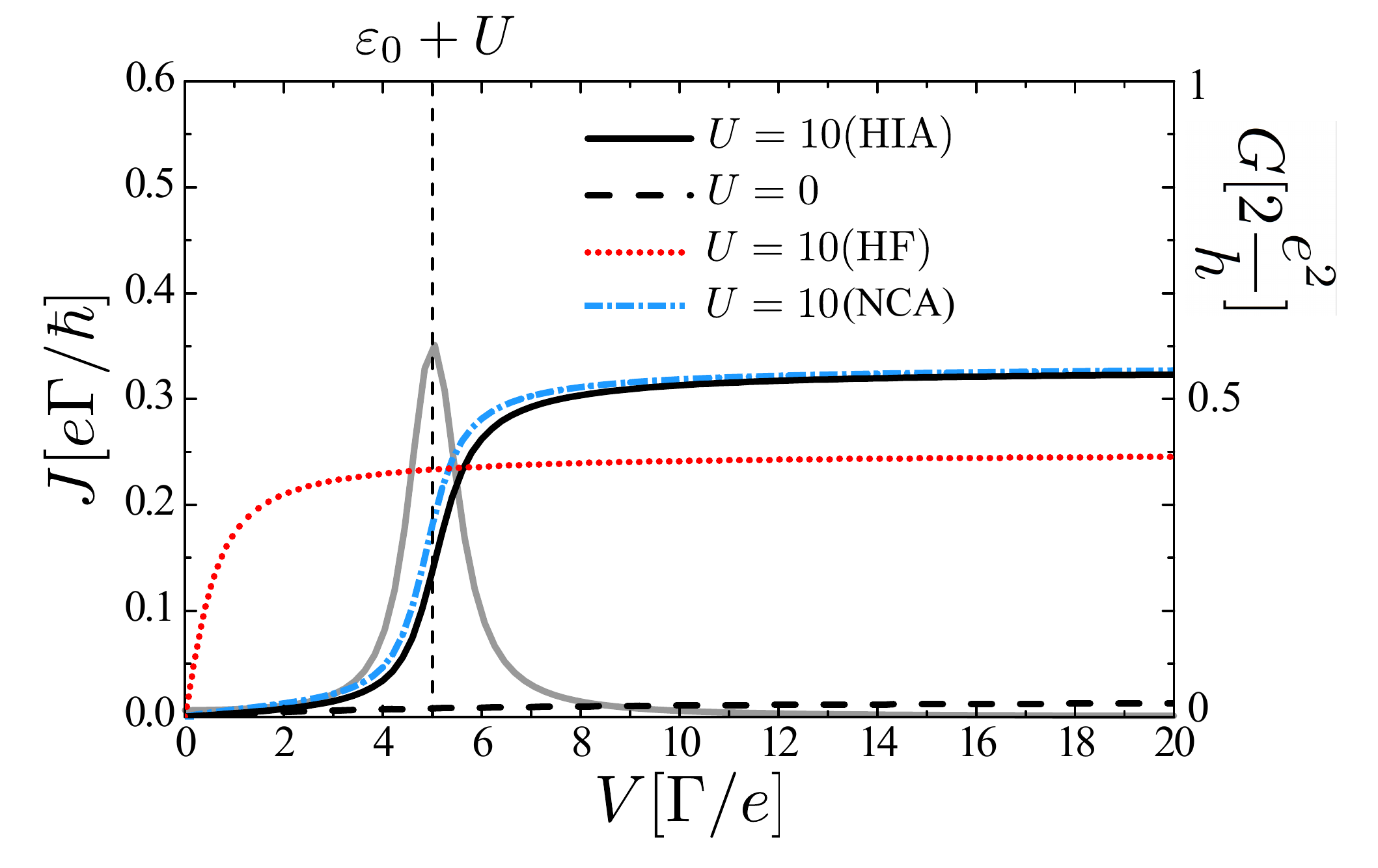}
        \end{center}
    \end{minipage}\\
    \caption{Same as Fig.\ref{pic:VolCur}, but with $\varepsilon_{0} = 0 $ (top) and $\varepsilon_{0}=-5$ (bottom).}
    \label{pic:VolCur2}
\end{figure}

In the wide band limit,
the real part of the retarded self-energy vanishes, while its imaginary part is constant
\begin{eqnarray}
\Lambda(\varepsilon) & = & 0\ , \\
\Gamma^{L(R)} (\varepsilon) & = & \Gamma^{L(R)}.
\end{eqnarray}
We can now analyse the current-voltage characteristic, with the voltage defined as $V = \mu_L - \mu_R$, and the symmetrized current $J = \frac{1}{2}(J_L - J_R)$ calculated from \eqref{eq:JLReq}.
Fig. \ref{pic:VolCur} depicts the calculation results for
the following parameters: $\mu_L = V,\,\mu_R = 0,\,\varepsilon_{0} = 5,\,U = 10,\,T=0.1,\,\Gamma_L=\Gamma_R=0.5$.
All energies are measured in $\Gamma$ units.
The results for $U=0$ and those obtained in Hartree-Fock
and noncrossing approximations
are also shown for comparison (this last one is discussed later). 
There are two jumps in $J(V)$ which
correspond to $V =\varepsilon_0$ and $V = \varepsilon_0 + U$.
One might think that $U$ lowers the current, as seen in Fig.~\ref{pic:VolCur}. However, it is not a general result:  the Coulomb repulsion can even raise it, as observed in the upper part  of Fig.~\ref{pic:VolCur2}  and more obviously in the lower part.
For the parameters used in these figures, a simple expression for the current, valid in the weak coupling limit ($ \Gamma \ll U, V$)
\begin{equation}
J = \frac {e \Gamma}{\hbar} \frac 1 2 \Bigl( (1- n_{\sigma}) (f_L -f_R)_{\varepsilon_0} + n_{\sigma} (f_L -f_R)_{\varepsilon_0+U} \Bigr),
\label{eq:Jweak}
\end{equation}
associated with a weak-coupling expression for density
$n_{\sigma} = (1- n_{\bar{\sigma}}) \displaystyle \bar{f} (\varepsilon_0) + n_{\bar{\sigma}} \displaystyle \bar{f} (\varepsilon_0+U)$,
enables to evaluate the current plateau values observed in the three plots.
The proximity between HIA and weak-coupling predictions for bias corresponding to the middle of the plateaus is smaller than 5\%.
However the transition between two consecutive plateaus is too abrupt in the weak-coupling approach due to its delta-shaped spectral weight.
$J$ increases with the number of conducting channels opened in the bias window, but not in proportion with this number. Indeed the Hubbard band spectral weight renormalizes each contribution, as explicitly  stated in Eq.  \eqref{eq:Jweak}.

To further validate the foundations of our
approach, that is HIA out of equilibrium, to which the formalism reduces under steady-state conditions,
we compare its predictions for the current with those evaluated within a more sophisticated approach, namely NCA. In NCA on-dot interaction is treated in a more reliable way than in HIA and the spectral density displays a more elaborate structure: showing at low temperature the Kondo resonance, and more
generally wider Hubbard bands (typically four times wider).
Thus for a quantitative comparison, we choose NCA parameters such as to obtain the same peak locations and bandwidths in both approaches~\cite{rem}.
As can be seen in Figs.~\ref{pic:VolCur} and \ref{pic:VolCur2}, we obtain an overall satisfying agreement  between NCA and HIA.
Even for the considered temperature $T = 0.1 \ \Gamma > T_K$, where $T_K$ is the Kondo temperature, some discrepancies can be observed, especially for $\epsilon_0 =0$, because of the incipient Kondo resonance. However
NCA slightly overestimates the Kondo resonance weight when this structure is close to $\epsilon_0$~\cite{WingreenMeir1994}.

\section{Time-dependent case}
In the time-dependent case it is necessary to solve Eqs. \eqref{eq:final1}-\eqref{eq:final2},
with $\Sigma^r(t_1,t_2)$ given by Eq. \eqref{eq:sigmar}.

\subsection{Wide-band limit}
It is possible to go further analytically
in the case of the wide-band limit, that is neglecting the influence of bandstructure details:
in that case $\rho (\varepsilon_k)$ is assumed to be independent of $\varepsilon_k$
and the couplings to the leads become $V_{k \alpha} (t) = u_{\alpha} (t) V_{\alpha}$ where $V_{\alpha}$
are constant. This leads to
$\Gamma^{L/R}(\varepsilon, t, t') = \Gamma^{L/R}(t, t')$,
and the retarded self-energy becomes
\begin{equation}
\Sigma^r (t, t')  =
-\frac{i}{2} \Gamma(t) \delta(t-t') ,
\end{equation}
where
$\Gamma(t) \equiv \sum_{\alpha} \Gamma^{\alpha}(t,t)  = 2 \pi \rho \sum_{\alpha}  u_{\alpha}^2(t)  |V_{\alpha}|^2 \equiv \sum_{\alpha}  \Gamma^{\alpha} u_{\alpha}^2(t) $.
It is then possible
to solve the equations for the retarded Green's functions
\begin{equation}
\tilde{G}_{\sigma\sigma,0(U)}^{r} (t,t')
= g_{0(U)}^r (t, t') \exp \left\{ -\int_{t'}^t \frac{\Gamma(t_1)}{2} d t_1 \right\}.
\label{eq:gwb}
\end{equation}
The expressions for $g_{0(U)}^r (t, t')$ were given previously in Eqs.~\eqref{eq:g1} and \eqref{eq:g2}.
The total retarded Green's function is then determined from Eq.~\eqref{eq:Grtottilde}.

It is worth noting that the analytical simple expression quoted in Eq. \eqref{eq:gwb} which is very convenient for numerical evaluation, is a direct consequence of the original manner used in this work to make the Hubbard I approximation.  With the canonical HIA, we do not obtain such a handy result.

The lesser Green's function can be evaluated using the Langreth
analytic continuation rules, see Eq. \eqref{eq:less12}.
To calculate the lesser Green's functions it is useful to define, following Ref~\cite{JauhoWingreenMeir1994}
\begin{eqnarray}
A_{L/R}^{0(U)} (\varepsilon, t) &=&  \int d t_1 u_{L/R} (t_1) \tilde{G}_{\sigma\sigma,0(U)}^{r} (t,t_1) e^{i \varepsilon (t-t_1)} \nonumber \\
&& \times\exp\left( -i\int_t^{t_1} d t_2 \Delta_{L/R} (t_2) \right) .
\label{eq:ALR1}
\end{eqnarray}
Using these functions it is possible to write the lesser Green's function in a compact form
\begin{eqnarray}
G_{\sigma \sigma}^< (t,t) & = & i n_{\sigma} (t) \nonumber \\
& = & i \sum_{L,R} \Gamma^{L/R} \int \frac{d \varepsilon}{2\pi} f_{L/R} (\varepsilon)
\left\{\left[1-n_{\bar{\sigma}}(t)\right] |A_{L/R}^{0}(\varepsilon, t)|^2 \right. \nonumber \\
& & + \left. n_{\bar{\sigma}}(t) |A_{L/R}^{U}(\varepsilon, t)|^2\right\}.
\end{eqnarray}
Similarly to the stationary case we have equation for $n_{\sigma} (t)$ which can be solved explicitly to yield
\begin{equation}
n_{\sigma} (t) = n_{\bar\sigma} (t) = \frac{n^0 (t)}{1+n^0 (t)-n^U (t)},
\label{eq:nwb}
\end{equation}
where
\begin{eqnarray}
n^{0(U)} (t) & = & \sum_{L,R} \Gamma^{L/R} \int \frac{d \varepsilon}{2\pi} \,
f_{L(R)} (\varepsilon) |A_{L/R}^{0(U)}(\varepsilon, t)|^2.
\end{eqnarray}
The current consists of two contributions $J_{L/R} (t) = J_{L/R}^{(1)} (t) + J_{L/R}^{(2)} (t)$, with
\begin{eqnarray}
J_{L/R}^{(1)} (t) & = & -\frac{e}{\hbar} \Gamma^{L/R}u_{L/R}^2(t) \left[ n_{\uparrow} (t) + n_{\downarrow} (t) \right], \\
\label{eq:JWB1}
J_{L/R}^{(2)} (t) & = & -\frac{e}{\hbar} \Gamma^{L/R} u_{L/R} (t) \sum_{\sigma} \int \frac{d \varepsilon}{\pi} f_{L/R} (\varepsilon) \nonumber \\
&& \times  \mathrm{Im} \left\{ B_{L/R}^{0}(\varepsilon, t) + B_{L/R}^{U}(\varepsilon, t)\right\}.
\label{eq:JWB2}
\end{eqnarray}
Quantities $B_{L/R}^{0(U)}$ appear after the integration of $G_{\sigma\sigma}^{r}$ over $t_1$ in the
general expression for the current \eqref{eq:lcur}. They can be written as
\begin{eqnarray}
B_{L/R}^{0} (\varepsilon, t) & = & \int d t_1 [1 - n_{\bar{\sigma}} (t_1)] u_{L/R} (t_1) \tilde{G}_{\sigma\sigma,0}^{r} (t,t_1) \nonumber \\
&& \times e^{i \varepsilon (t-t_1)}
\exp\left( -i\int_t^{t_1} d t_2 \Delta_{L/R} (t_2) \right), \nonumber \\
&& \\
B_{L/R}^{U} (\varepsilon, t) & = & \int d t_1 n_{\bar{\sigma}} (t_1) u_{L/R} (t_1) \tilde{G}_{\sigma\sigma,U}^{r} (t,t_1) \nonumber \\
&& \times e^{i \varepsilon (t-t_1)}
\exp\left( -i\int_t^{t_1} d t_2 \Delta_{L/R} (t_2) \right).\nonumber \\
&&
\end{eqnarray}
Details of the numerical procedure used for calculation of current for arbitrary time
dependences can be found in Appendix A.

\subsection{Pulse modulation}
\begin{figure}[h!]
\begin{minipage}[t]{.48\textwidth}
        \begin{center}
            \includegraphics[width=\textwidth]{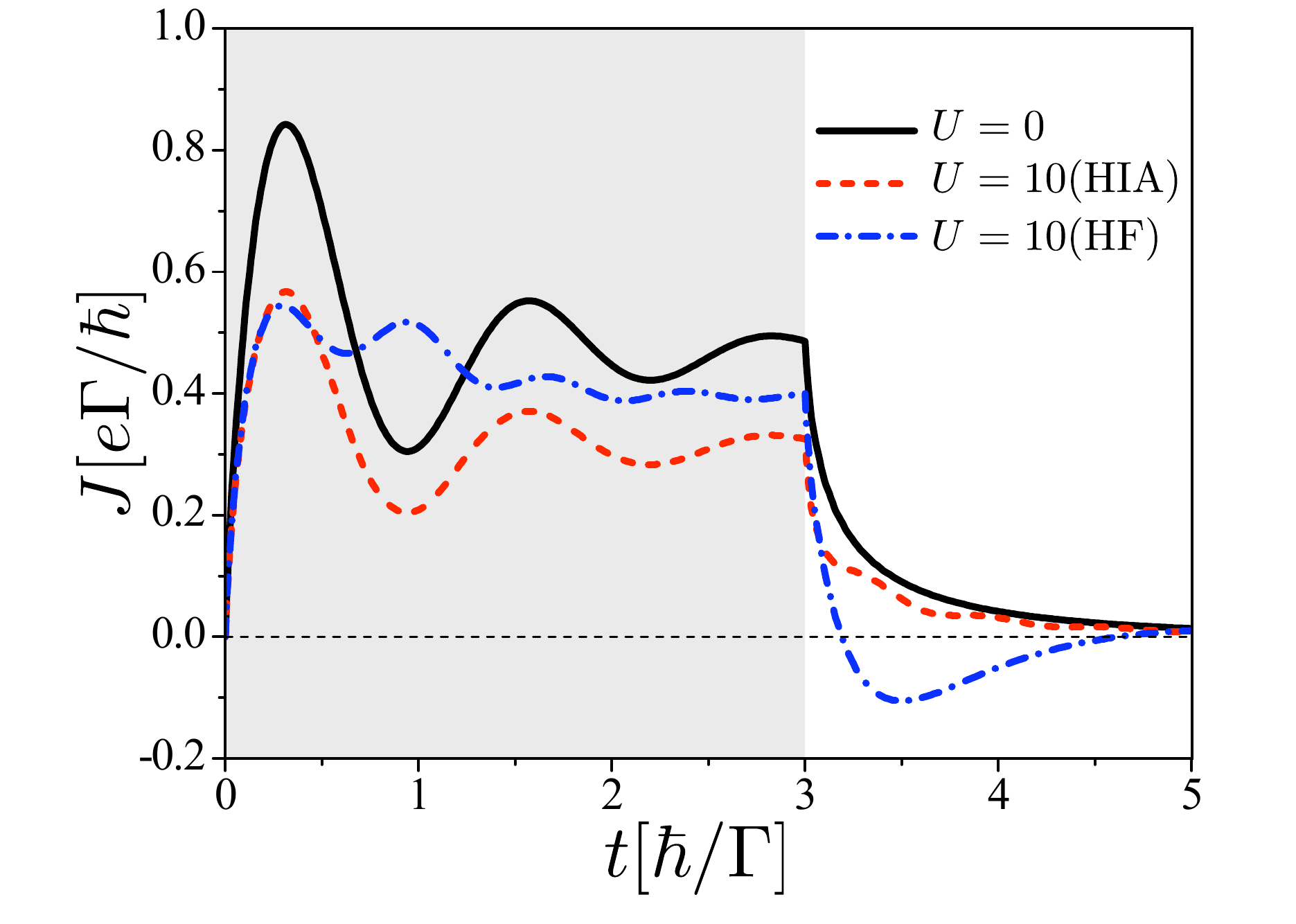}
        \end{center}
    \end{minipage}
    \begin{minipage}[t]{.44\textwidth}
        \begin{center}
            \includegraphics[width=\textwidth]{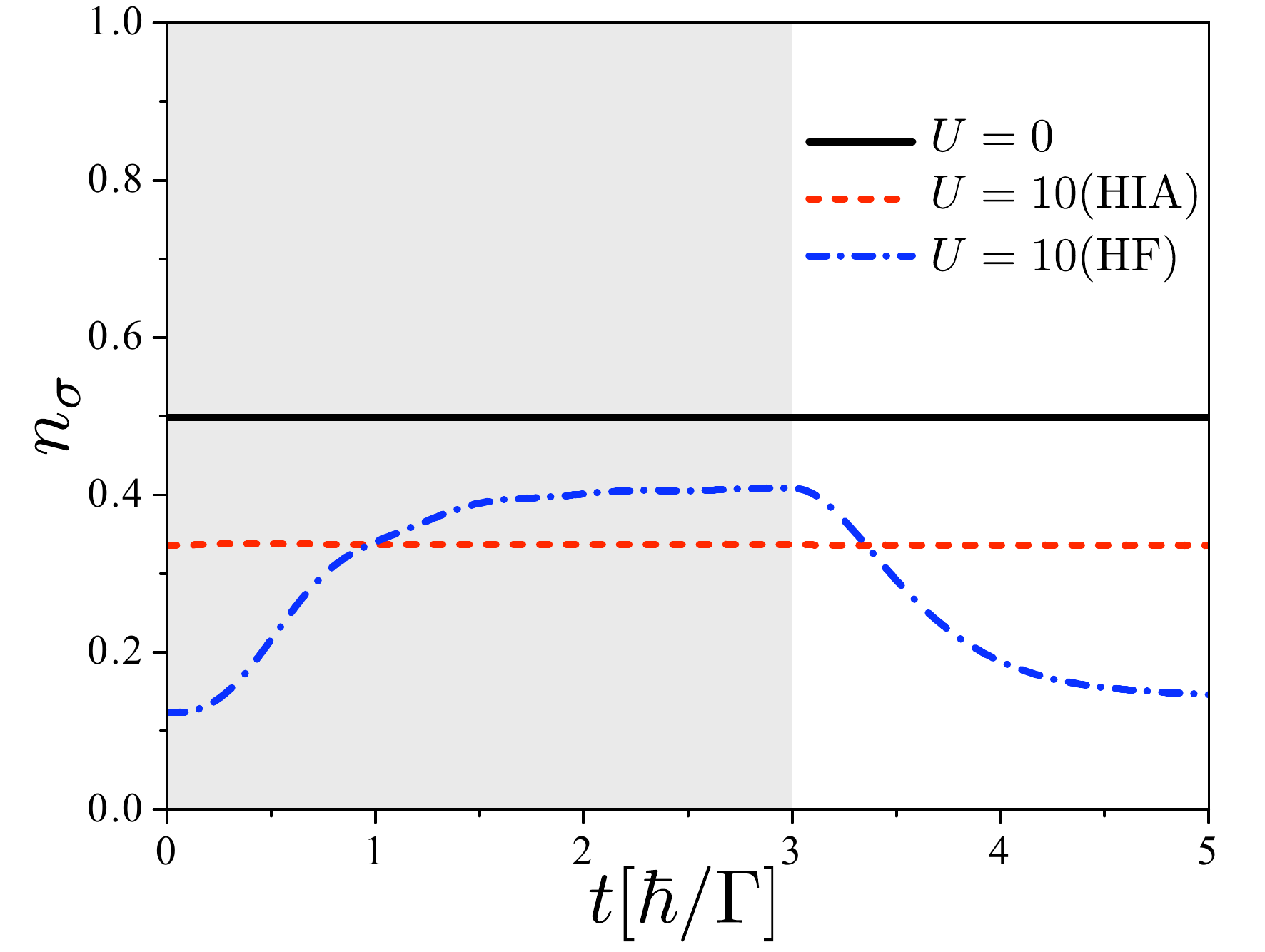}
        \end{center}
    \end{minipage}\\
    \caption{Time dependence of current (top) and occupancy per spin at the dot (bottom) in case
    of  pulse modulation with
    $\Gamma_L=\Gamma_R=0.5,\,T=0.1,\,\mu_L=0,\,\mu_R=0,\,\varepsilon_0=0,\,\Delta_L=10,\,\Delta_R=0,\Delta= 5$, $s=3$,
    for $U=0$ and $U=10$ (HF and HIA).
The gray area depicts the pulse duration.
}
    \label{pic:TimeDep1}
\end{figure}

In the case of a rectangular pulse shape modulation, we choose the following time dependences
\begin{eqnarray}
\Delta_{L/R} (t) & = & \left[\operatorname{\theta}(t) - \operatorname{\theta}(t-s)\right] \Delta_{L/R},
\label{eq:stepfirst}\\
\varepsilon_0 (t) & = & \varepsilon_0 + \left[\operatorname{\theta}(t) - \operatorname{\theta}(t-s)\right] \Delta,\\
u_{L/R}(t) & = & 1 ,
\end{eqnarray}
it entails that
\begin{eqnarray}
\Gamma^{L/R} (t) & = & \Gamma^{L/R} = \frac{1}{2}\Gamma,\\
\Sigma_{\sigma}(t) & = &-\frac{i}{2} \Gamma
\label{eq:steplast}.
\end{eqnarray}
Figure \ref{pic:TimeDep1} depicts
$J(t)$ and $n_{\sigma}(t)$ for the following choice of parameters:
$\Gamma_L=\Gamma_R=0.5,\,T=0.1,\,\mu_L=0,\,\mu_R=0,\,\varepsilon_0=0,\,\Delta_L=10,\,\Delta_R=0,\,\Delta=5$, $s=3$,
for $U=0$, and $U=10$ (HF and HIA), in $\Gamma$ unit for energy  and $\hbar/\Gamma$ unit for time.
For  finite $U$
in the HIA  the current behaves similarly
as in the case of no correlations,
experiencing ringing
(pseudo-oscillations)
 with the same period, but with reduced value
 when the transient regime fades. The HF approximation predicts a period which is roughly half as long.
A higher current value for $U=0$ than for $U=10$
can be explained as follows:
for $U=10$, only one channel corresponding to the
energy transition $\varepsilon_0+\Delta$
 lies in between $\mu_L+\Delta_L$ and $\mu_R+\Delta_R$ ($\varepsilon_0+ \Delta+U$  is outside the bias window) while for $U=0$ both channels contribute to the current.

The similarity between the non-interacting case and the HIA, as well as the discrepancy between these and the HF result is also obvious in the dot occupancy versus time plot (lower part of Fig. \ref{pic:TimeDep1}).
For $U=0$, $\epsilon_0(t)$ always lies in the middle of the bias window, then by symmetry $J_L(t) =-J_R(t)$, such that the dot occupancy is not affected by the bias step: $n_{\sigma} $ is therefore time-independent, and equals 1/2. The dot occupancy in the HIA is neither affected by the
bias onset for the same reason, and the obtained constant value of 1/3 can be understood on time-independent grounds.
This value attests the spectral weight transfer which takes place between the two Hubbard bands: indeed the weight of the lower Hubbard band is $1-n_{\bar{\sigma}} $, furthermore this band lies symmetrically around the middle of the bias window, for which  $\bar{f}(\epsilon) =1/2$, such that one has
$1/2 (1-n_{\bar{\sigma}} )  = n_{\sigma}$, hence leading to the value $n_{\sigma} = 1/3$.
The occupancy in HF approach at  $t=0$ is reduced compared to the non-interacting one due to the shift  of the band  towards higher energy.
Conversely and in an erroneous way, the HF result for $n_{\sigma}$ is time-dependent and
brings up an artificial characteristic time scale.
Quite generally charge conservation leads to
\begin{equation}
J_L(t)  + J_R (t) -e \sum_{\sigma} \frac{d n_{\sigma}}{dt}=0 \ ,
\end{equation}
where the first two terms are tunnel currents, while the last one is called displacement current.
Thus a time-independent dot occupancy is expected when, by symmetry,
$J_L(t) = -J_R (t)$.

\begin{figure}[h!]
\begin{minipage}[t]{.48\textwidth}
        \begin{center}
            \includegraphics[width=\textwidth]{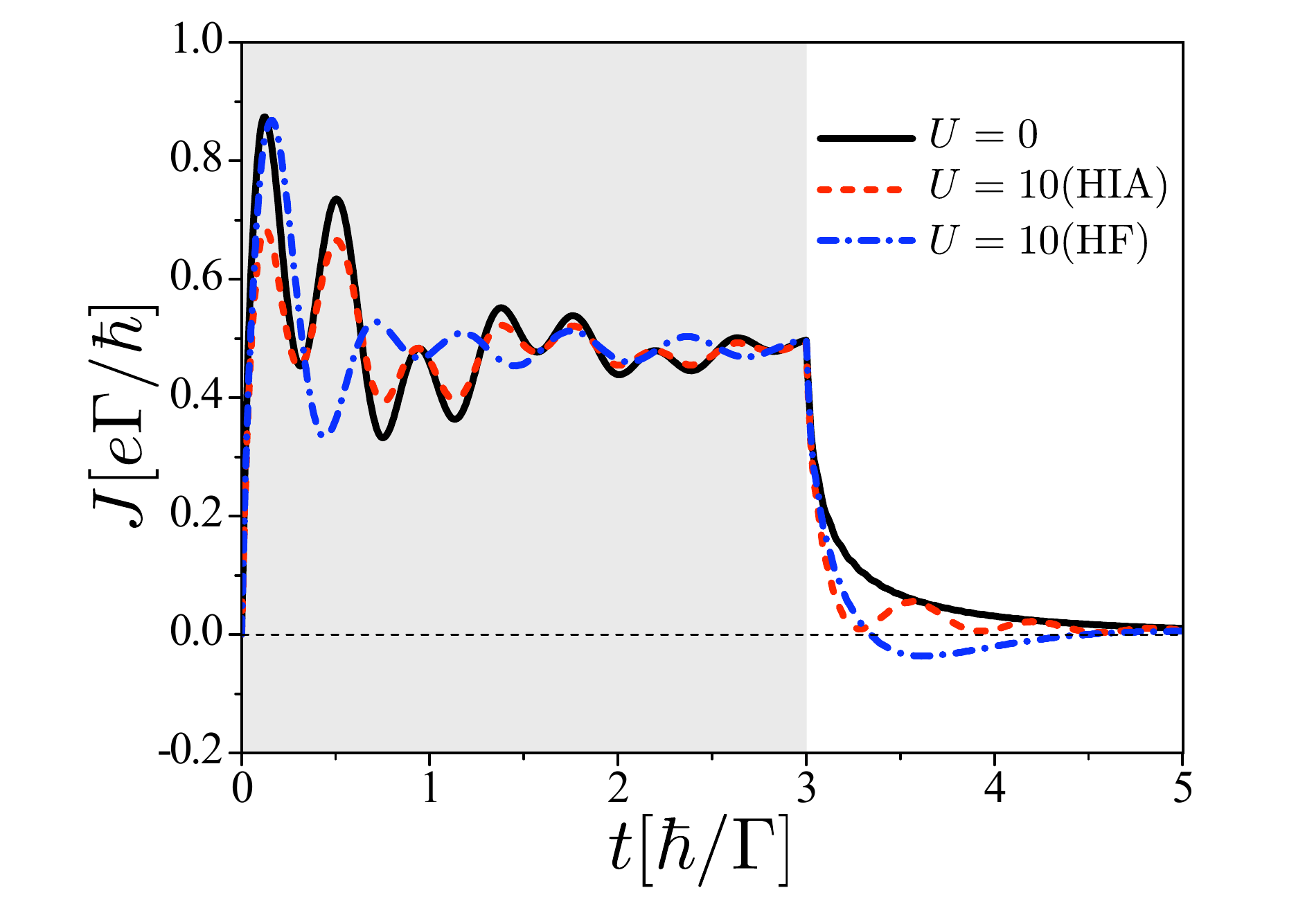}
        \end{center}
    \end{minipage}
    \begin{minipage}[t]{.44\textwidth}
        \begin{center}
            \includegraphics[width=\textwidth]{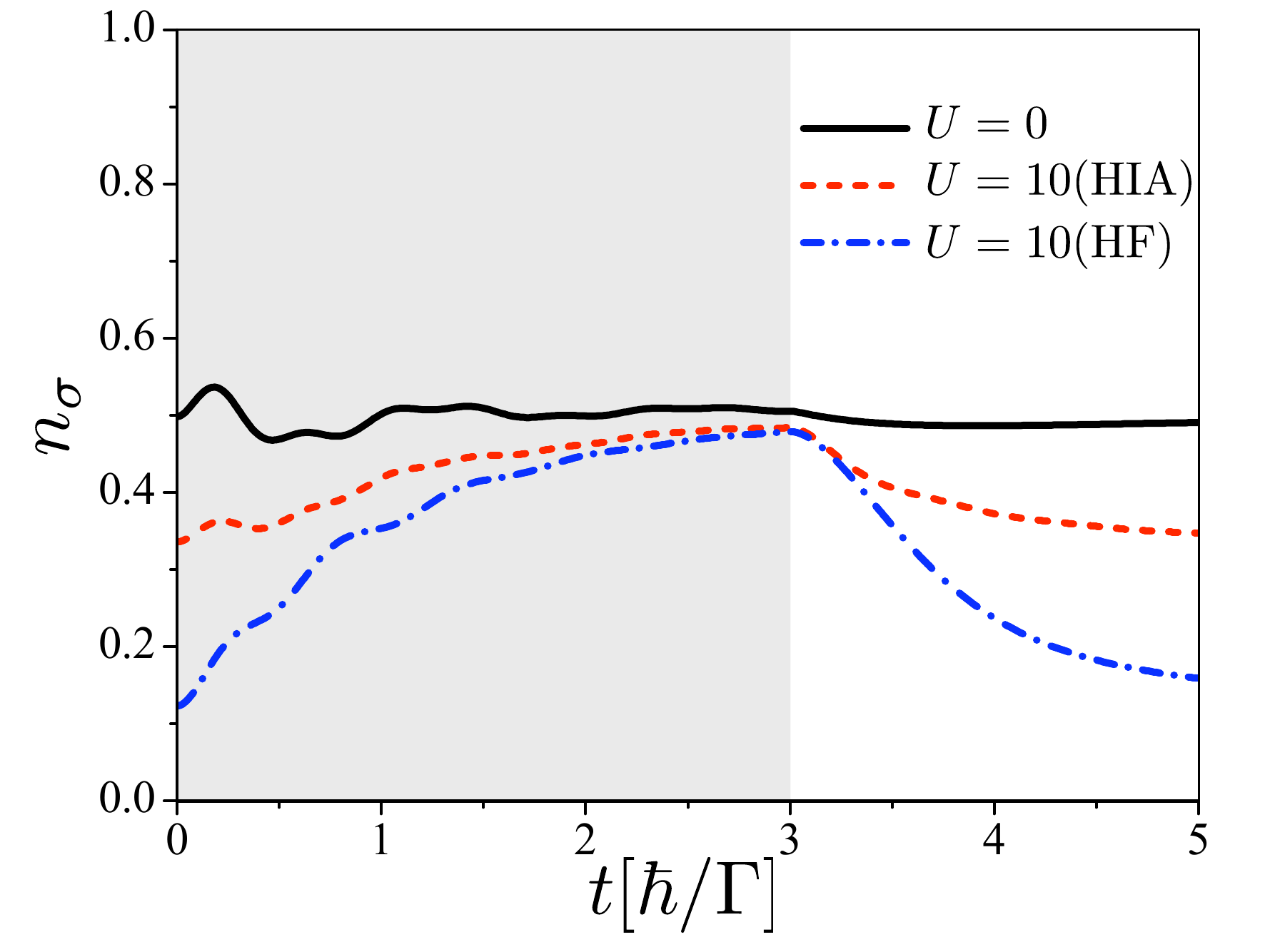}
        \end{center}
    \end{minipage}\\
    \caption{Same as Fig.\ref{pic:TimeDep1}, but for $\Delta_L=20$.}
    \label{pic:TimeDep2}
\end{figure}

\begin{figure}[h!]
\begin{minipage}[t]{.44\textwidth}
        \begin{center}
            \includegraphics[width=\textwidth]{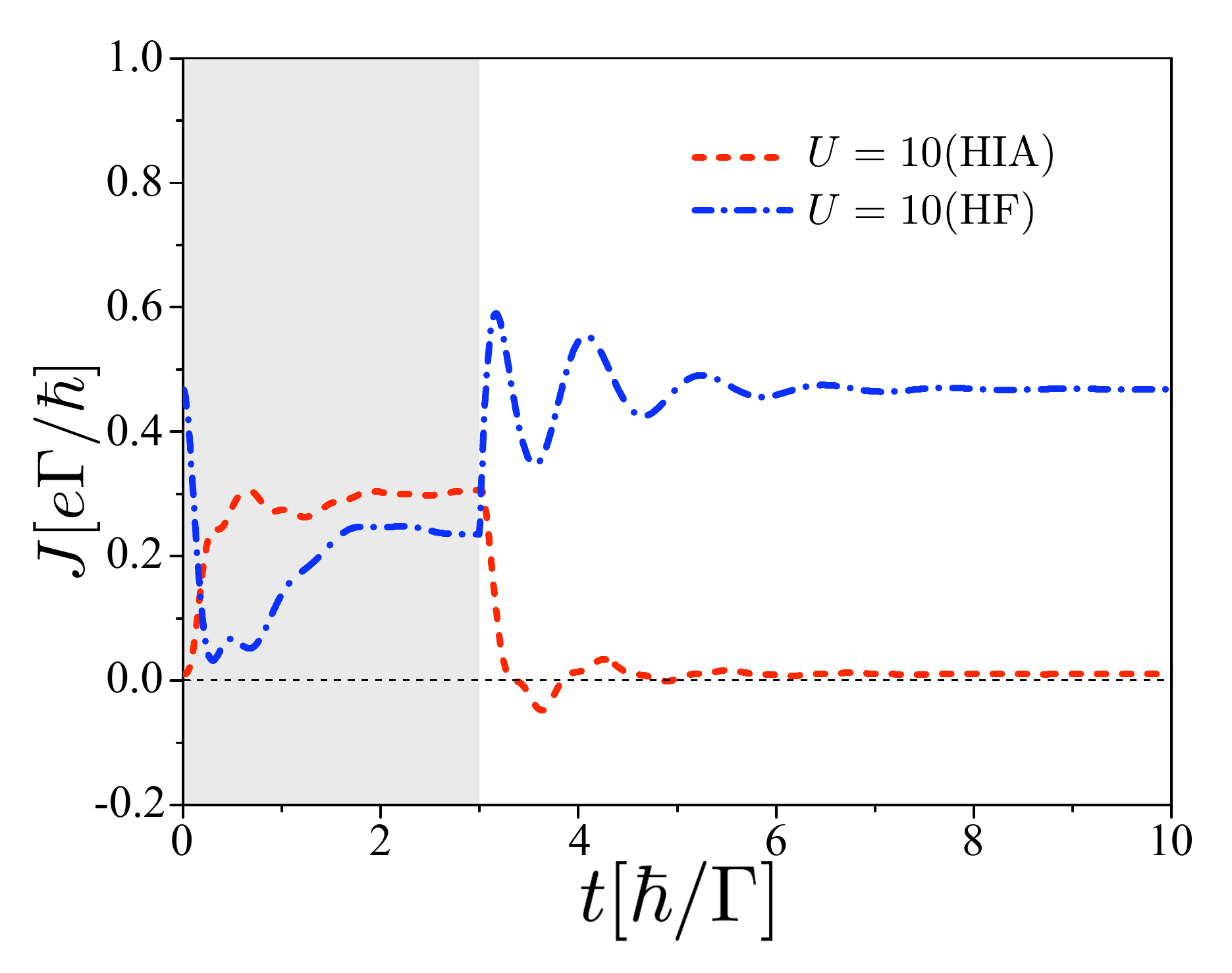}
        \end{center}
    \end{minipage}
    \begin{minipage}[t]{.44\textwidth}
        \begin{center}
            \includegraphics[width=\textwidth]{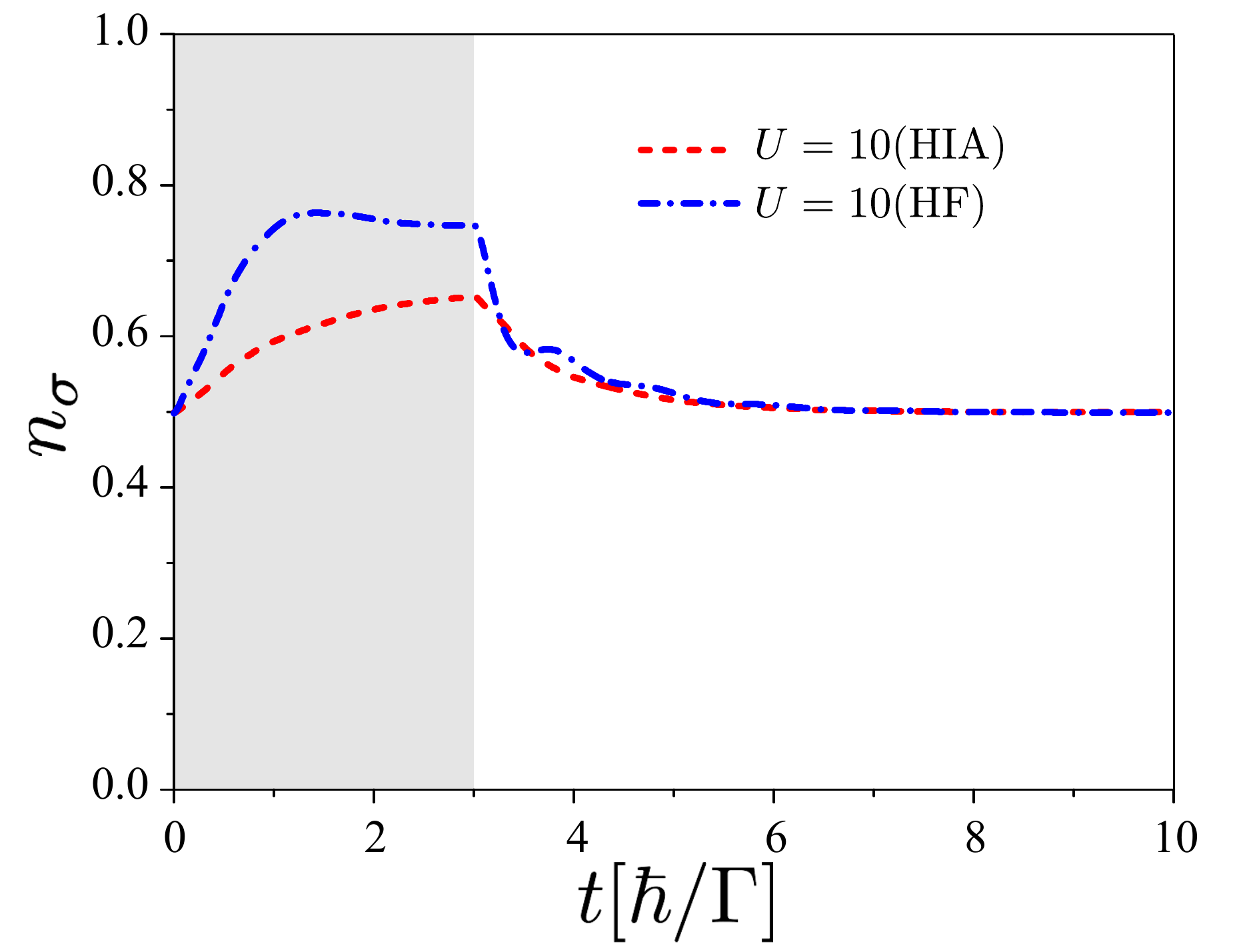}
        \end{center}
    \end{minipage}\\
    \caption{Same as Fig.\ref{pic:TimeDep1}, but for $\mu_L = 10\ , \varepsilon_0 = -5\ , \Delta_L=0 \ , \Delta =-10$.}
    \label{pic:TimeDep3}
\end{figure}

When both bands lie in the bias window, as in Fig. \ref{pic:TimeDep2}, the differences for current and occupancy between $U=0$ and HIA for $U=10$  are attenuated.
As previously noted, the periods of pseudo-oscillations of $J(t)$ are nearly the same between $U=0$ and HIA, while the amplitudes are slightly different.
Now the density is affected by the bias setup, even for $U=0$, due to an asymmetric distribution of spectral weight in the bias window.
All these results, for one or two bands inside the bias window,
display a transient regime which differs depending on whether the
bias is turned on or off.
The greatest qualitative difference between non-interacting and HIA results occurs during the equilibrium restoration.
We observe that the over-current values are quite close in HF and HIA in Fig.  \ref{pic:TimeDep1}, but it may be
fortuitous: it is not the case in Fig. \ref{pic:TimeDep2}.
The HF has an  additional shortcoming, predicting a temporary sign reversal of the current immediately after the pulse end, a behavior absent in the HIA and non interacting cases, which can be attributed in part to an overestimation of the dot occupancy in the steady state regime combined with an underestimation of $n_{\sigma}$ in the equilibrium regime.
Finally we choose the voltages and local dot parameters in such a way as to visualize the Coulomb blockade - conducting transition.
This is shown in Fig.~\ref{pic:TimeDep3} where, at $t=0$  the dot leaves the insulating Coulomb blockade region to enter the conducting one until $t=s$.
Letting $s \rightarrow \infty$ enables to access the charging time of the dot: for the present parameters, using an exponential modelization, we find $\tau \sim 1.2\ \hbar /\Gamma$ in HIA, this is about twice as long as the time predicted by HF, as seen in Fig.~ \ref{pic:TimeDep3}.

\begin{figure}[h!]
   \begin{center}
     \includegraphics[width=.48\textwidth]{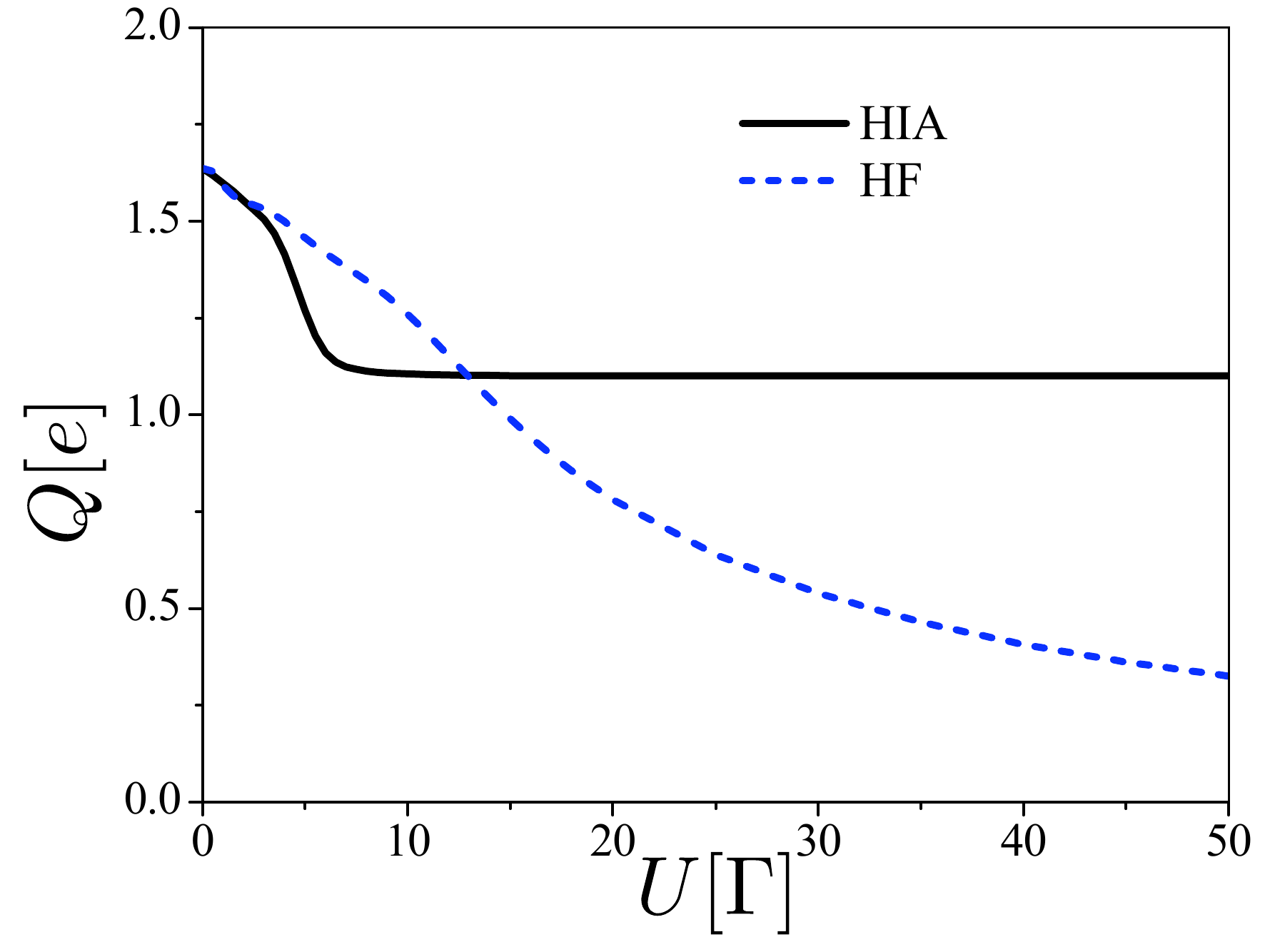}
     \caption{
    Transferred charge  per pulse
     versus $U$ in HF and HIA, for $\mu_L = \mu_R =0, \ \varepsilon_0=0, \ T=0.1, \ \Delta_L =10, \ \Delta_R =0, \ \Delta =5, \ s=3$.}
     \label{pic:ChargeU}
   \end{center}
\end{figure}

To explore further the discrepancies between HF and HIA, we analyze the dependence on $U$ of the transferred charge or
time-integrated symmetrized current.
The interval between consecutive pulses is assumed to be much larger than the length of the pulse, therefore
we can treat consecutive pulses independently.
Since the length of the pulse $s$ is less than time duration of the transient regime
(see Figs.~\ref{pic:TimeDep1}-\ref{pic:TimeDep2}),
this charge $Q$ can illustrate properties
of purely time-dependent phenomena, it is shown in Fig.~\ref{pic:ChargeU}.
{%
For small $U$, both HF and HIA give the same result, as expected since they converge to the exact description for $U=0$.
In the HIA, increasing $U$ from 0, the  transferred charge $Q$
decreases to settle at a constant value
when the higher Hubbard band leaves the bias window.}
The value $U=5$ which corresponds to $\varepsilon_0+\Delta +U = \mu_L+\Delta_L$, marks the upturn between $U=0$ and $U \rightarrow \infty$ regimes. The smoothness of this decrease depends on $\Gamma$.
Besides, in the HF approximation, the charge
transferred $Q$
keeps decreasing with the increase of $U$.
The Hubbard I approximation is known to contain more physics than HF approximation,
e.g. in stationary
and
equilibrium cases, and we can conclude that HF
approximation is also insufficient to describe time-dependent transport
in presence of Coulomb repulsion
\cite{myohaetal}.

\begin{figure}[h!]
   \begin{center}
     \includegraphics[width=.5\textwidth]{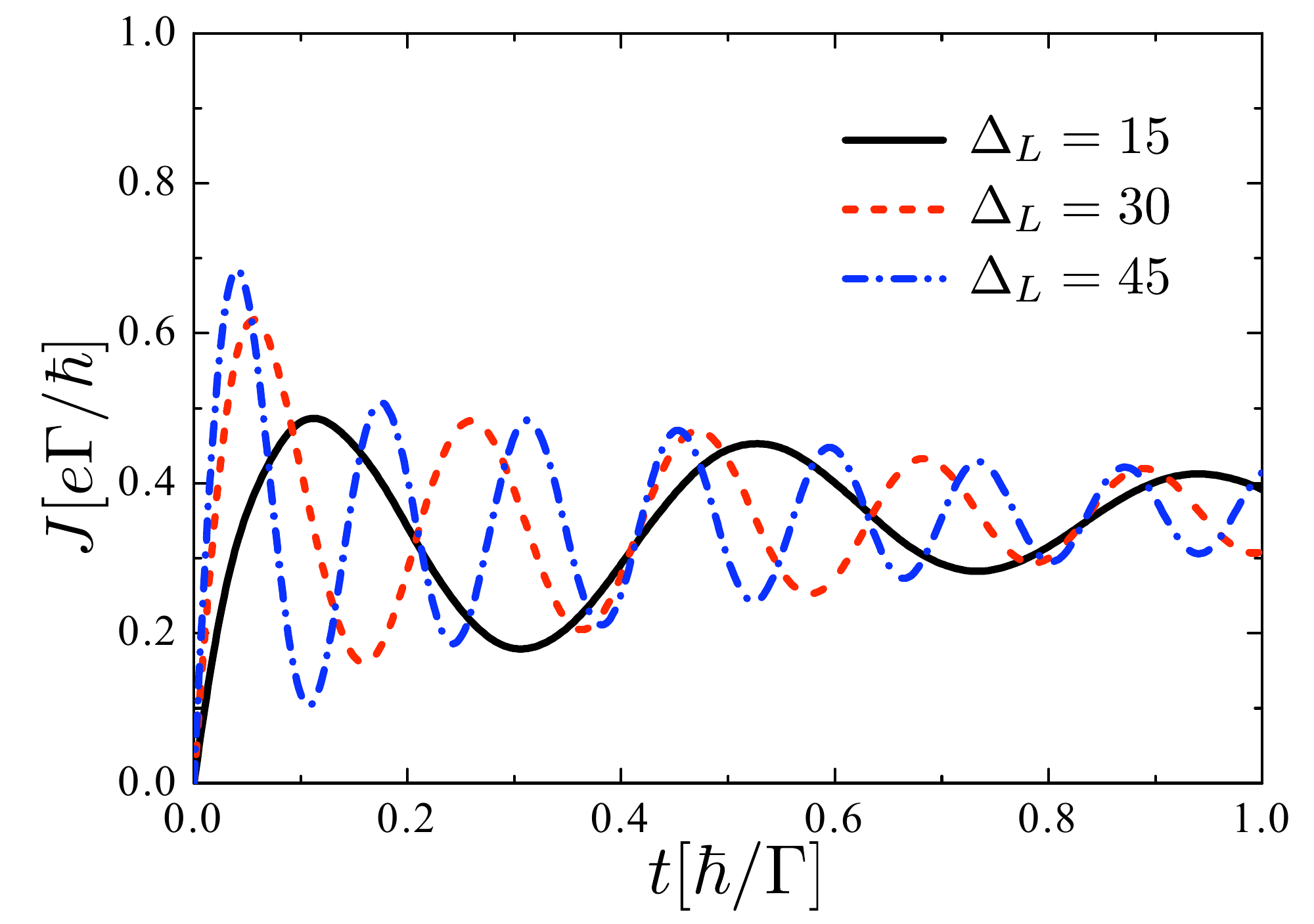}
     \caption{$J(t)$ in case of step voltage
     for different values of $\Delta_L$, for $\mu_L = \mu_R =0, \ \varepsilon_0 =0, \ U=10, \ T=0.1, \ \Delta_R = \Delta =0, \ s\ge 1$.}
     \label{pic:Period}
   \end{center}
\end{figure}

It is interesting to explore which parameters can influence the period of the ``ringing''
in the time-dependence
of $J(t)$.
When we change all the parameters $\Delta_L$, $\Delta_R$ and $\Delta$, the time-dependence can be rather
complicated (see Fig. \ref{pic:TimeDep2}), so
we can focus on only one parameter.
Let us fix the values of $\Delta_R$ and $\Delta$ to zero and change only $\Delta_L$.
Figure \ref{pic:Period} depicts $J(t)$ in case of
a step-function characterized by
$\Gamma_L=\Gamma_R=0.5,\,T=0.1,\,\mu_L=0,\,\mu_R=0,\varepsilon_0=0,\,U=10, s \to \infty$,
and with $\Delta_L =15, 30, 45$.
The value of the steady-state
current does not change with $\Delta_L$:
indeed in all cases the conducting channels are the same.
In the meantime, the period is strongly affected by $\Delta_L$:
the product of $\Delta_L$ and period $T_0$
approximately satisfies
$\Delta_L \cdot T_0  \sim 2 \pi$.
This can be attributed to the presence of a
phase multiplier of the form  $\exp(i \Delta(t') t')$
in the expressions which determine the time dependence of the current.
This result is compatible with the observation reported in Ref~\cite{myohanenepl098} where the oscillations are ascribed to the electronic transitions between the lower dot state and the leads potentials.

\begin{figure}[h!]
   \begin{center}
     \includegraphics[width=.5\textwidth]{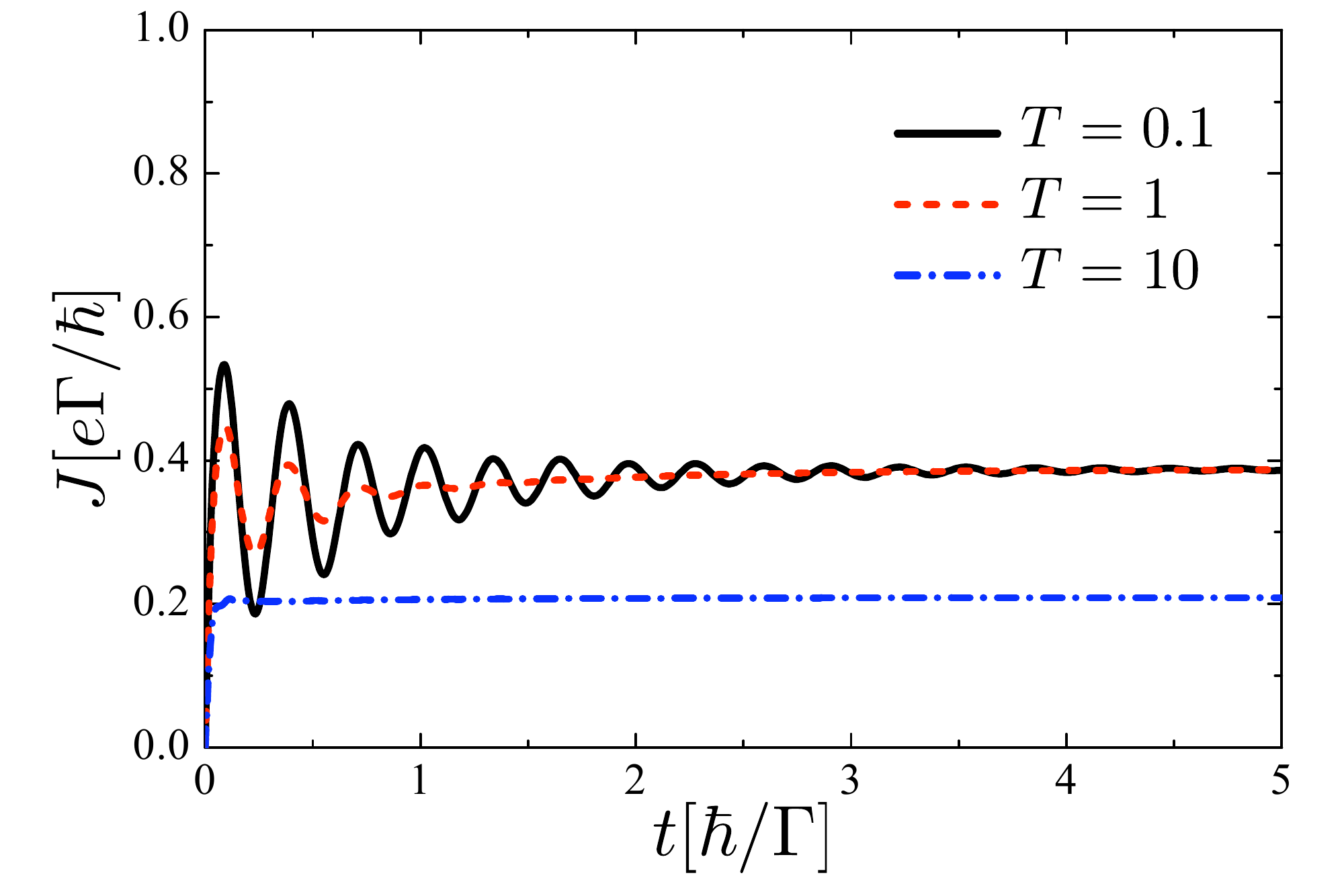}
     \caption{$J(t)$ in case of step voltage
     for different temperatures, for  $\mu_L = \mu_R =0, \ \varepsilon_0 =0, \ U=10, \ \Delta_L=20, \ \Delta_R = \Delta =0, \ s\ge 5$.}
     \label{pic:Temp}
   \end{center}
\end{figure}

Finally we can also study the temperature influence on $J(t)$.
Figure \ref{pic:Temp} depicts $J(t)$ for different
$T$ values.
Moderate temperature has little influence on ``ringing'' period,
but can change significantly the amplitude of these oscillations.
Besides, very high temperature affects
the steady-state value of the current, which
is attained  almost instantaneously. So, it
is seen that complex time-dependence is
restricted to low temperatures.

\subsection{Harmonic modulation}

\begin{figure}[h!]
\begin{minipage}[t]{.48\textwidth}
        \begin{center}
            \includegraphics[width=\textwidth]{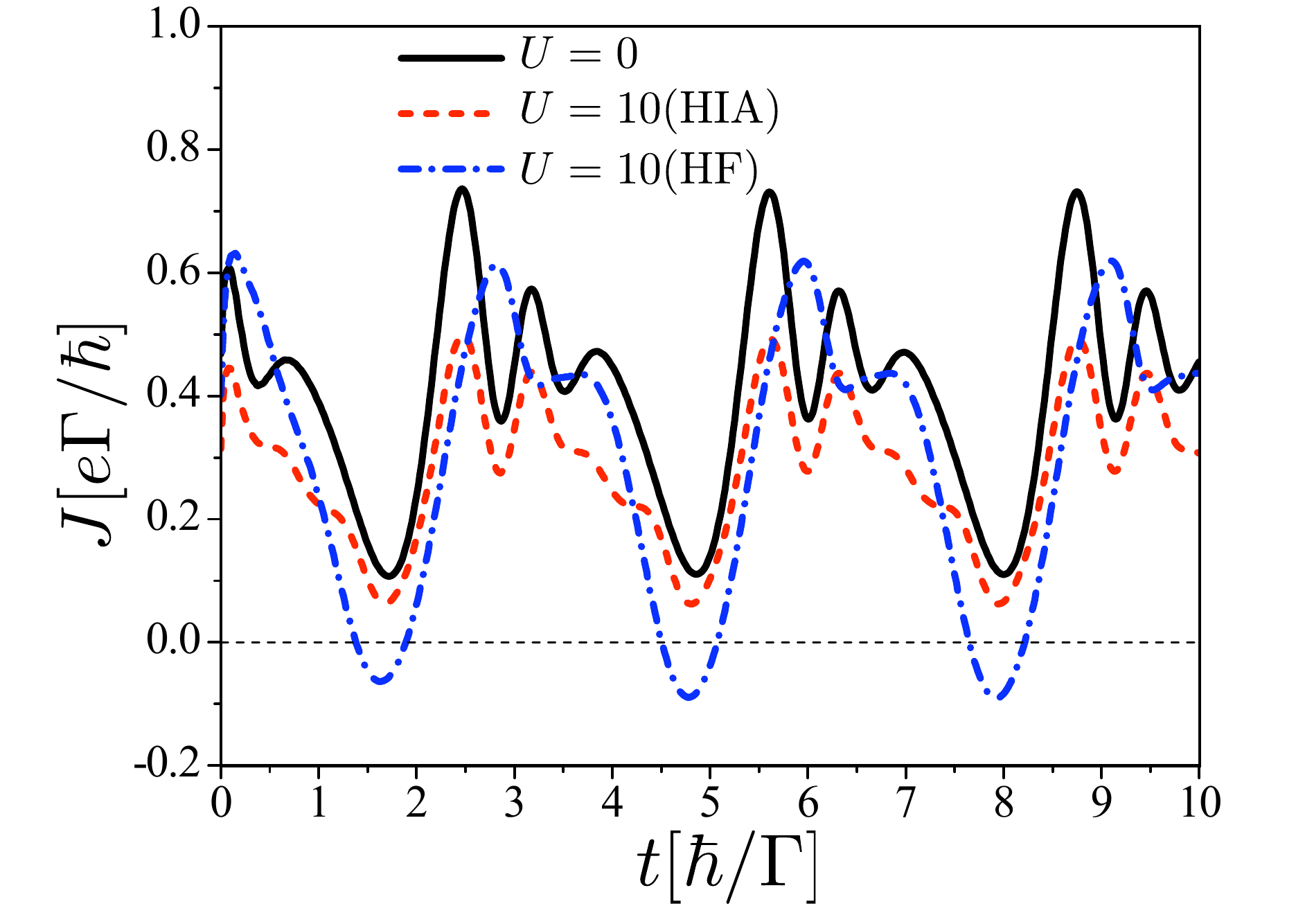}
        \end{center}
    \end{minipage}
    \begin{minipage}[t]{.48\textwidth}
        \begin{center}
            \includegraphics[width=\textwidth]{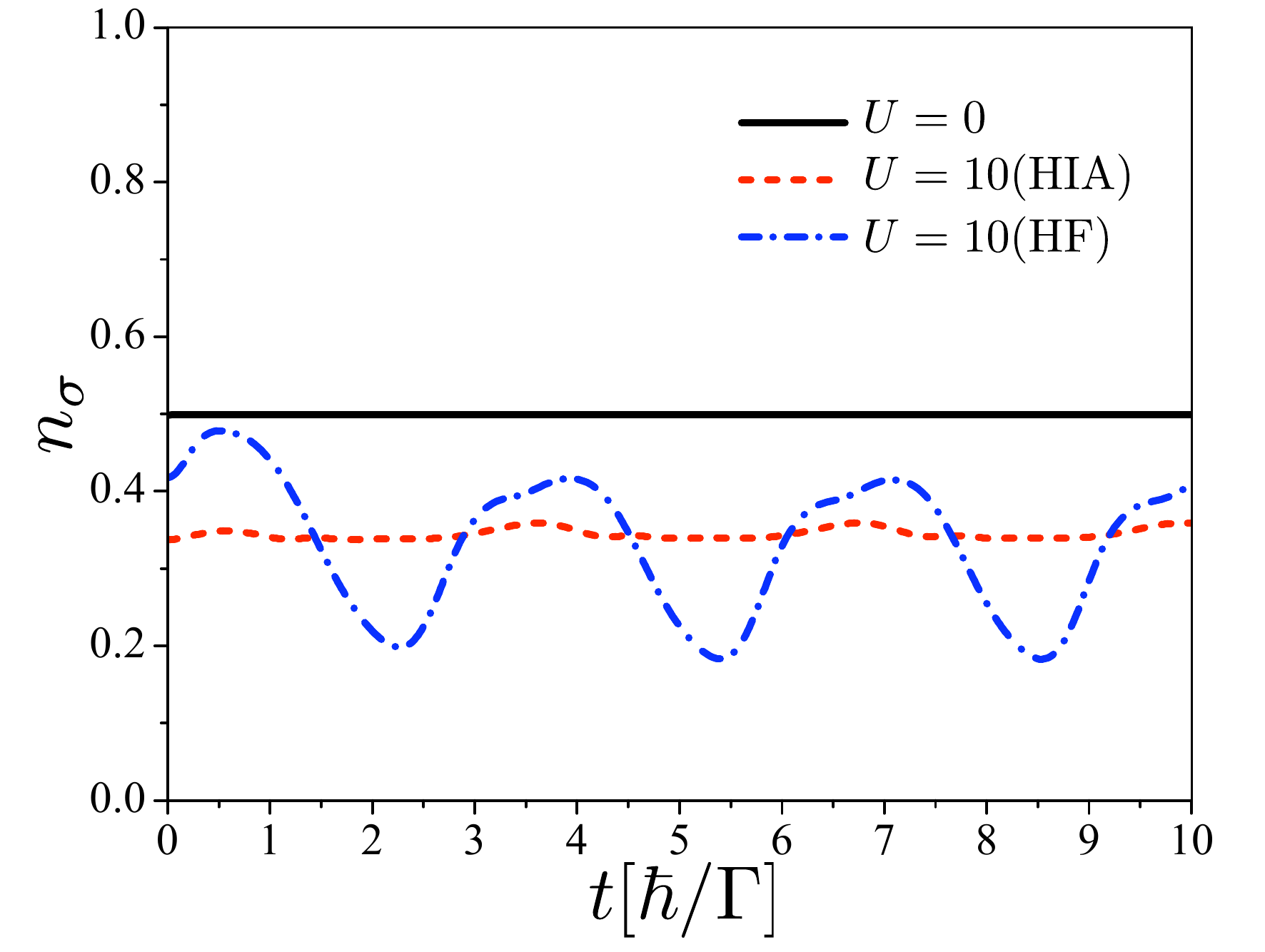}
        \end{center}
    \end{minipage}\\
    \caption{Time dependence of current (top) and dot  occupancy  per spin (bottom) in case
    of harmonic modulation, for
    $\Gamma_L=\Gamma_R=0.5,\,T=0.1,\,\mu_L=10,\,\mu_R=0,\,\varepsilon_0=5,\,\Delta_L=10,\,\Delta_R=0,\Delta= 5$, $\omega=2$,
    for $U=0$ (exact solution) and $U=10$.}
    \label{pic:TimeDepHarm}
\end{figure}

It is interesting to explore the case when external voltages are periodic in time
and how correlations, after the transient regimes, influence the forced ones.
In case of harmonic modulation we choose
in-phase voltages:
$\Delta_{L/R,0} (t) = \Delta_{L/R,0} \cos(\omega t)$.
This kind of modulation was studied before for the non-interacting case in Ref.~\cite{JauhoWingreenMeir1994}
and in Ref.~\cite{CroySaalmann86} with an exponential modulation of the hybridization;
it was generalized to the interacting model in the HF approximation~\cite{Deus2012}.
Such an harmonic time-dependence was also adressed in the Kondo regime~\cite{Lopez1998}\cite{Arrachea2008}.
Figure \ref{pic:TimeDepHarm} depicts the time dependence of the current (top) and dot  occupancy
 per spin (bottom) in case of harmonic modulation, for
the following parameters:
$\Gamma_L=\Gamma_R=0.5,\,T=0.1,\,\mu_L=10,\,\mu_R=0,\,\varepsilon_0=5,\,\Delta_L=10,\,\Delta_R=0,\Delta=5$, $\omega=2$ (pulsation measured in $\Gamma / \hbar$),
for $U=0$ and $U=10$ (HF and HIA).

In HIA, the modulation amplitude for dot occupancy is very slight, but not strictly zero: it stays close to 1/3, and never reaches 2/5, which would be expected once per period, if the system were adiabatic.
Eventually, in this forced regime, the current becomes periodic in time, with the same period as
the external perturbation.
For the chosen parameters, a
rise and fall
regime sets in:
the higher
Hubbard band
alternatively reaches the border and leaves the varying bias window.
However a higher frequency (about three times larger) also emerges.

As also previously argued, the HF approximation is not reliable in the transient regime, not more in the forced regime, predicting a periodic  current inversion, as well as a poor estimation of dot occupancy.

\subsection{Adiabatic and non-adiabatic pumping}

\begin{figure}[h!]
   \begin{center}
     \includegraphics[width=.44\textwidth]{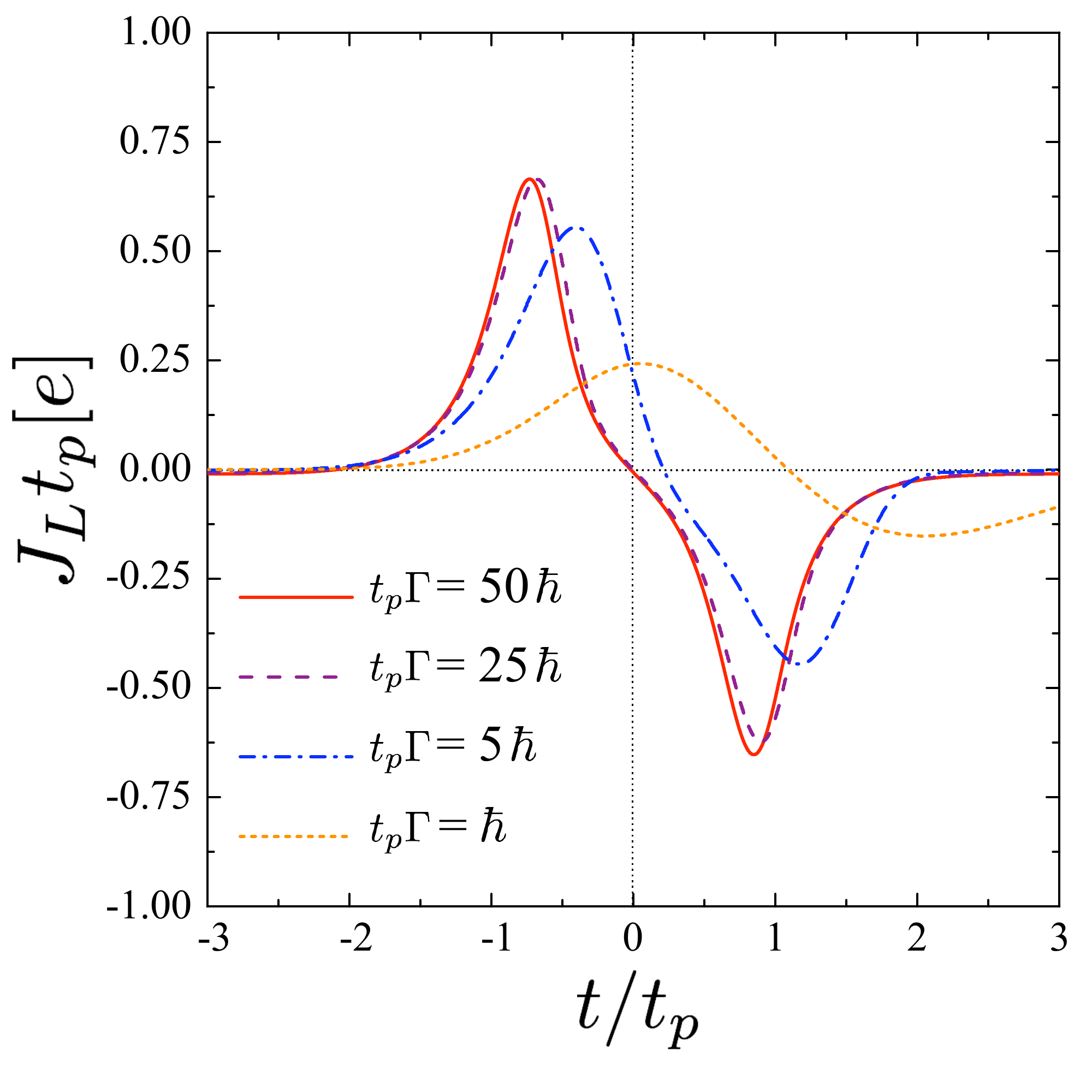}
     \caption{
     Left current versus time for a gaussian-time-dependent gate voltage $\varepsilon_{0}(t) = 1-2 e^{-(t/t_p)^2}$,  in the HIA for different values of $t_p$, the other parameters are $\mu_R = \mu_L =0,\,U = 10,\,T=0.1,\,\Gamma_L=\Gamma_R=0.5$.
     }
     \label{pic:gaussian-gate}
   \end{center}
\end{figure}

As previously mentioned in the harmonic regime, current and densities do not follow adiabatic predictions, even at low frequency $\hbar \omega \le \Gamma$~\cite{Cavaliere09}. The non-adiabatic behavior is exploited in charge pumping.
The idea of producing a current  in the absence of any bias voltage, called pure pumping, can also be addressed within the HIA.
A recent paper focussed on this issue in a formalism close, but not equivalent, to the present work~\cite{CroySaalmann85}.
These authors truncate the hierarchy of equation of motion at the same level than detailed above, and manage also to circumvent double-time
Green's functions evaluation, using an auxiliary-mode expansion.
It thus seemed to us interesting to make a detailed comparison between these two resembling techniques.
In the footsteps of these authors, we looked at the case of a gaussian pulse gate voltage, and calculated the left current. Results are shown in Fig.~\ref{pic:gaussian-gate}. Varying the gaussian time-width $t_p$ allows to browse the cases of adiabatic and non adiabatic response.
There is a qualitative agreement, and a very good quantitative consistency for rapid pulse (low $t_p$); however some discrepancies in the adiabatic region are observed, the current being higher in our HIA. The origin of this difference in not very clear for us but stems from different correlation treatment procedures; indeed for $U=0$
our numerical procedure for arbitrary time dependence was checked by a detailed comparison with Jauho's et al. results~\cite{JauhoWingreenMeir1994},  we also agree with uncorrelated Croy {\it et al.}'s results.

\section{Conclusions}
A time-dependent formalism for a single
level
interacting quantum dot coupled
to two leads has been developed within the Hubbard I approximation,
in a convenient and handy way
for numerical evaluation.
It enables us to consider general time dependences for hybridization as well as for gate or bias voltages.
To validate the approach,
the steady-state regime has been, in some parameter range, favorably compared with NCA results.

The formalism results
in the appearance of two Hubbard bands
in the local spectral density.
For those bands we introduce two Green's functions, which offer the key advantage to get rid of double-time evaluation, without any a priori assumption about adiabatic or sudden limits.
These bands undergo spectral weight transfers which influence the transport properties, in contrast with the rigid band frame.

Calculations
show the influence of Coulomb correlations
on the current, which mainly consists in a change of amplitude.
The influence on the time structure (e.g. ``ringing'' period), appear
to be mostly insignificant.
Comparisons between Hubbard I and Hartree-Fock
approximations show that the latter is insufficient to describe time-dependent transport.

The presented method can also be extended beyond the wide band limit,
indeed the key point consists in writing the dot Green's function as a sum of two independent Green's functions; this acts upstream, before the assumption of wide band limit. To implement our formalism beyond this limit deserves further study.

\section*{Acknowledgements}
We thank Steffen Sch\"{a}fer and Oleh Fedkevych
for valuable discussions. V.V. appreciates hospitality of staff
at IM2NP in Marseilles, where part of this work was done and
acknowledges support from the
Ministry of Education and Science of Ukraine
(Program ``100+100+100'' for studying and interning abroad).
D.A. was supported by the program ``Microscopical and phenomenological models of fundamental physical
processes at micro and macro scales'' (Section of physics and astronomy of the NAS of Ukraine).

\section*{Appendix A: Numerical procedure}

Here we describe the numerical procedure used for
calculating the current and the dot occupancy for arbitrary
time dependences in the wide-band limit for the HIA.
The time dependences are $\Delta_{L(R)}(t)$, $\varepsilon_0 (t) = \varepsilon_0 + \Delta(t)$,
$u_{L(R)}(t)$ with
the only condition being that all time-dependent perturbations start at $t_0$. For the sake of simplicity we choose
in this appendix $t_0=0$, i.e.
for $t<0$ we have a stationary state. The time-dependent current and dot occupancy are calculated
by integrating the $A_{L(R)}^{0(U)} (\varepsilon, t)$ and $B_{L(R)}^{0(U)} (\varepsilon, t)$ functions.
The key point here is the numerical computation of these functions.

We will split the integration, e.g. in the expression \eqref{eq:ALR1} for $A_{L(R)}^{0(U)}$
into two parts: $\int_{-\infty}^t = \int_{-\infty}^0 + \int_{0}^t$. Then, after performing
the integration in the first term we get
\begin{eqnarray}
&& A_{L/R}^{0(U)} (\varepsilon, t)  =  \frac{\exp\left[i(\varepsilon-\varepsilon_{0(U)})t\right]}{\varepsilon-\varepsilon_{0(U)}+\frac{i\Gamma}{2}} \nonumber \\
&& \times 
\exp\left[-i\int_0^{t} dt_1 \left(\Delta(t_1)-\Delta_{L(R)}(t_1) - \frac{i\Gamma(t_1)}{2}\right) \right]  \nonumber \\
& & - i \int_{0}^t d t_1 \exp\left[i \varepsilon (t-t_1) \right] u_{L/R}(t_1) \exp\left[ -i \int_{t_1}^t d t_2 \varepsilon_{0(U)} (t_2) \right]  \nonumber \\
& & \times  \exp\left[ i \int_{t_1}^t d t_2 \Delta_{L/R} (t_2) \right] \exp\left[- \int_{t_1}^t  \frac{\Gamma(t')}{2} d t' \right].
\label{eq:A1WB}
\end{eqnarray}
Expressions for $B_{L/R}^{0(U)}$ are also expressed in this form, for instance we get
\begin{eqnarray}
&& B_{L/R}^{0} (\varepsilon, t) = \frac{(1-n_{\bar\sigma}^0)\exp\left[i(\varepsilon-\varepsilon_0)t\right]}{\varepsilon-\varepsilon_0+\frac{i\Gamma}{2}} \nonumber \\
&& \exp\left[-i\int_0^{t} dt_1 \left(\Delta(t_1)-\Delta_{L(R)}(t_1) - \frac{i\Gamma(t_1)}{2}\right) \right]  \nonumber \\
& & - i \int_{0}^t d t_1 \left(1-n_{\bar\sigma}(t_1)\right) \exp\left[i \varepsilon (t-t_1) \right] u_{L/R}(t_1) \nonumber \\
&& \times \exp\left[ -i \int_{t_1}^t d t_2 \varepsilon_0 (t_2) \right]   \exp\left[ i \int_{t_1}^t d t_2 \Delta_{L/R} (t_2) \right] \nonumber \\
& & \times  \exp\left[- \int_{t_1}^t  \frac{\Gamma(t')}{2} d t' \right],
\end{eqnarray}
where $n_{\bar\sigma}^0$ is the stationary value prior to modulation.

At $t=0$ we can compute the stationary
current and occupancy. We have
\begin{eqnarray}
A_{L/R}^{0(U)} (\varepsilon, 0) & = & \frac{1}{\varepsilon-\varepsilon_{0(U)}+\frac{i\Gamma}{2}}, \\
B_{L/R}^{0} (\varepsilon, 0) & = & \frac{(1-n_{\bar\sigma}^0)}{\varepsilon-\varepsilon_0+\frac{i\Gamma}{2}}, \\
B_{L/R}^{U} (\varepsilon, 0) & = & \frac{n_{\bar\sigma}^0}{\varepsilon-\varepsilon_0-U+\frac{i\Gamma}{2}},
\end{eqnarray}
from which we can compute $n_{\sigma}^0$ and then $J^0$, using Eqs. \eqref{eq:nwb}-\eqref{eq:JWB2}.

To compute transport properties for $t>0$, we express $A_{L/R}^{0(U)} (\varepsilon, t+dt)$ using Eq. \eqref{eq:A1WB}
 in terms of   $A_{L/R}^{0(U)} (\varepsilon, t)$ at previous time step:

\begin{eqnarray}
&&A_{L/R}^{0(U)} (\varepsilon, t+dt)  =  A_{L/R}^{0(U)} (\varepsilon, t) \exp\left(i(\varepsilon-\varepsilon_{0(U)})dt\right) \times \nonumber \\
&& \exp\left[-i\int_t^{t+dt} dt_1 \left(\Delta(t_1)-\Delta_{L(R)}(t_1) - \frac{i\Gamma(t_1)}{2}\right) \right]  \nonumber \\
& & - i \int_{t}^{t+dt} d t_1 \exp\left[i \varepsilon (t+dt-t_1) \right] u_{L/R}(t_1) \times \nonumber \\
&&  \exp\left[ -i \int_{t_1}^{t+dt} d t_2 \varepsilon_{0(U)} (t_2) \right] \times \exp\left[ i \int_{t_1}^{t+dt} d t_2 \Delta_{L/R} (t_2) \right]  \nonumber \\
& &  \exp\left[- \int_{t_1}^{t+dt}  \frac{\Gamma(t')}{2} d t' \right].
\end{eqnarray}
Because $dt$ is small we can use the midpoint method to perform integration from $t$ to $t+dt$
\begin{eqnarray}
&& A_{L/R}^{0(U)} (\varepsilon, t+dt)  \approx  A_{L/R}^{0(U)} (\varepsilon, t) \exp\left(i(\varepsilon-\varepsilon_{0(U)})dt\right)  \nonumber \\
&& \times \exp\left[-i dt \left(\Delta(\tilde{t})-\Delta_{L(R)}(\tilde{t}) - \frac{i\Gamma(\tilde{t})}{2}\right) \right]  \nonumber \\
& & - i dt \exp\left[i \varepsilon \frac{dt}{2} \right] u_{L/R}(\tilde{t})  \nonumber \\
&&\times  \exp\left[ -i \frac{dt}{2} \left( \varepsilon_{0(U)} (\tilde{\tilde{t}}) - \Delta_{L/R} (\tilde{\tilde{t}}) - \frac{i\Gamma(\tilde{\tilde{t}})}{2}\right)\right],
\end{eqnarray}
where $\tilde{t} = t + \frac{dt}{2}$ and $\tilde{\tilde{t}} = t + \frac{3dt}{4}$. 

For $B_{L/R}^{0(U)}$ it is possible to obtain similar approximation but $n_{\bar\sigma}$ also appears  in the
integration from $t$ to $t+dt$. We will use
$n_{\bar\sigma} \left(t + \frac{dt}{2}\right) \approx \displaystyle{ \frac{1}{2} \left(n_{\bar\sigma} (t) + n_{\bar\sigma} \left(t + dt\right)\right) }$.
Indeed we can compute $n_{\bar\sigma} \left(t + dt\right)$ before computing $B_{L/R}^{0(U)} (\varepsilon, t+dt)$
because it is determined only by $A_{L/R}^{0(U)}$. In the end, we get, e.g. for $B_{L/R}^{0}$
\begin{eqnarray}
&& B_{L/R}^{0} (\varepsilon, t+dt)  \approx  B_{L/R}^{0} (\varepsilon, t) \exp\left(i(\varepsilon-\varepsilon_0)dt\right) \nonumber \\
&&  \times \exp\left[-i dt \left(\Delta(\tilde{t})-\Delta_{L(R)}(\tilde{t}) - \frac{i\Gamma(\tilde{t})}{2}\right) \right]  \nonumber \\
& & - i dt \frac{2 - n_{\bar\sigma} (t) - n_{\bar\sigma} \left(t + dt\right)}{2} \exp\left[i \varepsilon \frac{dt}{2} \right] u_{L/R}(\tilde{t}) \nonumber \\
& &   \times \exp\left[ -i \frac{dt}{2} \left( \varepsilon_0 (\tilde{\tilde{t}}) - \Delta_{L/R} (\tilde{\tilde{t}}) - \frac{i\Gamma(\tilde{\tilde{t}})}{2}\right)\right].
\end{eqnarray}

Computing and integrating the central quantities $A_{L(R)}^{0(U)} (\varepsilon, t)$ and $B_{L(R)}^{0(U)} (\varepsilon, t)$ in the previously presented calculations, typically require less than one minute on a 1.7 GHz Core 2 Duo.


\begin{thebibliography}{99}
\bibitem{rem} To obtain similar bandwidths in NCA and HIA, we choose $\Gamma$ four times smaller in NCA.
\end{thebibliography}
\end{document}